\documentclass[apj]{emulateapj}
\usepackage{apjfonts}
\usepackage{epsf}
\newcommand{\Msun}{\mathrm{M}_{\odot}}
\newcommand{\Lsun}{\mathrm{L}_{\odot}}

\newcommand{\bdv}[1]{\mbox{\boldmath$#1$}}
\def\ba{{\bdv{a}}}
\def\bx{{\bdv{x}}}
\def\bsigma{{\bdv{\sigma}}}
\def\bdelta{{\bdv{\Delta}}}

\begin{document}

\slugcomment{Submitted to ApJ}
\shortauthors{Gnedin et al.}
\shorttitle{Tully-Fisher Relation}

\title{Dark Matter Halos of Disk Galaxies: \\
       Constraints from the Tully-Fisher Relation}
        
\author{ Oleg Y. Gnedin\altaffilmark{1},
         David H. Weinberg\altaffilmark{1},
         James Pizagno\altaffilmark{1},\\
         Francisco Prada\altaffilmark{2},
         Hans-Walter Rix\altaffilmark{3}}

\altaffiltext{1}{The Ohio State University,
    Department of Astronomy,
    140 W 18th Ave., Columbus, OH 43210;
    \mbox{\tt ognedin@astronomy.ohio-state.edu}}
\altaffiltext{2}{Ramon y Cajal Fellow, 
    Instituto de Astrofisica de Andalucia (CSIC), 
    E-18008 Granada, Spain}
\altaffiltext{3}{Max-Planck-Institute for Astronomy, 
    Konigst\"uhl 17, D-69117, 
    Heidelberg, Germany}

\begin{abstract}
We investigate structural properties of dark matter halos of disk
galaxies in hierarchical $\Lambda$CDM cosmology, using a well-defined
sample of 81 disk-dominated galaxies from the SDSS redshift survey.
We model the mass-velocity (TF) and fundamental plane (FP) relations
of these galaxies, which are constructed from the galaxy stellar mass,
disk scale length, and optical H$\alpha$ rotation velocity at 2.2
scale lengths.  We calculate a sequence of model galaxy populations,
defined by the distribution of the stellar disk-to-total mass
fraction, $m_d$.  We include the effect of adiabatic contraction of
dark matter halos in response to condensation of baryons.  We find
that models with constant $m_d$ under-predict the intrinsic scatter of
the TF and FP relations and predict an (unobserved) strong correlation
between TF residuals, even with the full range of halo concentration
scatter.  Introducing a scatter of disk mass fractions and allowing
the mean value $\bar{m}_d$ to scale with the stellar surface density
significantly improves observational match of both the slope and
intercept of the model TF relation and reduces the predicted residual
correlation enough to be statistically consistent with the data.  The
distribution of angular momentum parameters $\lambda_d$ required to
match the observed disk scale lengths is significantly narrower than
that predicted for halo spin parameters.  Our best-fit models with a
Kroupa stellar IMF over-produce the galaxy stellar mass function and
predict the virial $r$-band mass-to-light ratios, $M_{\rm vir}/L_r$,
systematically lower than those inferred from galaxy-galaxy weak
lensing and satellite dynamics.  We investigate three possible
solutions to these problems: (1) ignoring the effects of adiabatic
contraction, (2) adopting a ``light'' stellar IMF with $M_*/L$ lower
than the Kroupa IMF by 0.15 dex, or (3) considering the lower halo
concentrations predicted for a low cosmological power spectrum
normalization $\sigma_8 \approx 0.74$.  In combination with our
proposed correlation of $\bar{m}_d$ with stellar surface density, any
of these solutions yields acceptable residual correlations and
relieves most of the observational tension between the TF relation and
the galaxy stellar mass function.
\end{abstract}


\section{Introduction}
  \label{sec:intro}

In the standard theoretical framework of galaxy formation, baryons
cool, condense, and form stars in the centers of dark matter halos
\citep{white_rees78}.  Observed disk galaxies obey a tight correlation
between luminosity and rotation speed, known as the Tully-Fisher
(hereafter TF) relation \citep{tully_fisher77}.  The slope, intercept,
and scatter of the TF relation are critical constraints on galaxy
formation models \citep[e.g.,][]{cole_kaiser89, kauffmann_etal93,
cole_etal94, eisenstein_loeb96, somerville_primack99,
steinmetz_navarro99}.  These constraints can be characterized clearly
within the dissipative collapse modeling framework developed by
\citet{fall_efstathiou80} and \citet{gunn83}, updated to the cold dark
matter (CDM) scenario by \citet{dalcanton_etal97} and
\citet{mo_etal98}.  In this framework, the disk rotation velocity is
determined by the ratio, $m_d$, of the disk mass to the total halo
mass, by the halo density profile, and by the angular momentum
parameter, $\lambda_d$, which sets the disk scale length.  In analogy
to ``fundamental plane'' studies of the elliptical galaxy population
\citep{djorgovski_davis87, dressler_etal87}, one can consider disk
scale length as an additional parameter in galaxy scaling relations
\citep{shen_etal02}.  Since $V^2 \sim GM(R)/R$, ``maximal disk''
models in which baryons dominate the observed rotation curve predict a
strong anti-correlation between TF residual and disk size
\citep{courteau_rix99}.

In this paper, we derive empirical constraints on the distributions of
$m_d$ and $\lambda_d$ by modeling the sample of disk-dominated
galaxies from \citet{pizagno_etal05}.  This sample is comprised of 81
galaxies with H$\alpha$ rotation curves selected from the Sloan
Digital Sky Survey \citep[SDSS;][]{york_etal00} main galaxy redshift
sample \citep{strauss_etal02} in the absolute magnitude range $-18 >
M_{r} > -23$, for which disk-bulge decomposition yields a best-fit
$i$-band bulge fraction $\leq 0.1$.  \citet{pizagno_etal05} estimate
galaxy stellar masses $M_*$ from the observed luminosities and colors,
using the population synthesis models of \citet{bell_etal03} to relate
the color to the mean stellar mass-to-light ratio.  They characterize
rotation velocities by the amplitude of the observed rotation curve
at 2.2 disk scale lengths, $V_{2.2}$.  We construct model galaxy
populations with different $P(m_d)$ distributions and test their
ability to reproduce the parameters of the TF relation and its
residual correlations, after imposing the observed $M_* - R_d$
distribution as a constraint.  Our approach differs from previous
studies by using a new, homogeneous data set with well-defined
selection criteria and small observational errors, by working with
estimated stellar masses instead of luminosities, and by deriving the
$P(\lambda_d)$ distribution empirically from the data instead of
imposing it {\it a priori} from theory.  While the distribution of the
{\it halo} angular momentum parameter $P(\lambda)$ has been well
studied with $N$-body simulations \citep[e.g.,][]{barnes_efstathiou87,
bullock_etal01b}, the distribution of disk angular momenta could be
different because baryons and dark matter exchange angular momentum,
because a biased subset of halo baryons settle into the disk, or
because disk-dominated galaxies form in a biased subset of dark matter
halos.  An empirical determination of $P(\lambda_d)$ is therefore a
valuable diagnostic of galaxy formation physics.

For our purposes, we define a model of the disk galaxy population by
the probability distribution of $m_d$, which may depend on stellar
mass and scale length, $P(m_d | M_*,R_d)$.  We assume that the density
distribution of dark matter halos is given by the NFW profile
\citep{navarro_etal97} with a concentration parameter $c$.  The
distribution function $P(c)$ and mass dependence of the halo
concentrations have been carefully studied and are no longer a
systematic uncertainty of the models, but they influence the scatter
and slope of the predicted TF relation.  For each observed galaxy, we
draw random values of $m_d$ and $c$ consistent with the assumed
$P(m_d)$ and theoretically estimated $P(c)$, and calculate the halo
mass as $M = M_*/m_d$.  Note that our definition of $m_d$ includes
only the stellar mass and {\it does not} include the contribution of
cold gas in the disk.  We make this choice of necessity because we do
not have gas mass measurements; in the range of our sample, gas
fractions are typically $\sim 15\%$.  We determine the disk angular
momentum parameter $\lambda_d$ of each galaxy by matching the observed
disk scale length $R_d$.

We judge the acceptability of a model by how well it reproduces the
observed joint distributions of $M_*, R_d$, and $V_{2.2}$.  Since we
use the values of stellar masses and disk sizes as input to the
models, the bivariate $M_* - R_d$ distribution is reproduced by
construction.  The slope, amplitude, and scatter of the $M_* -
V_{2.2}$ relation and the correlation of residuals of the $M_* - R_d$
and $M_* - V_{2.2}$ relations ($\Delta R_d$ and $\Delta V_{2.2}$,
respectively) serve as tests.  Alternatively, we consider a
``fundamental plane''-type (FP) relation for all three variables,
$M_*$, $R_d$, and $V_{2.2}$.  After finding the distributions $P(m_d |
M_*,R_d)$ that give acceptable fits to the observed relations, we
examine the ability of these models to fit the observed baryon mass
function of galaxies.  We also make predictions for extended mass
distributions around the late-type galaxies, which are measured by
weak lensing and satellite dynamics.

We include in our models the effect of adiabatic contraction (AC) of
dark matter halos in response to central condensation of baryons.  The
deeper potential well created by the cooling and compression of gas
attracts more dark matter to the central regions of the galaxy and
increases its density.  In the analytic literature on disk galaxy
modeling, this effect has sometimes been omitted, or its inclusion
treated as a ``free parameter'', on the suspicion that the spherical
symmetry and adiabatic growth assumptions used in the standard
derivation of \citet{blumenthal_etal86} might make the results invalid
in the hierarchical merging paradigm of galaxy formation.  Recently,
the AC effect has been tested and quantified for galaxies forming by
hierarchical merging, using ultrahigh resolution gasdynamics
cosmological simulations with cooling and star formation
\citep{gnedin_etal04}.  The effect is not as strong as predicted by
the original \citet{blumenthal_etal86} model, mainly because the dark
matter particle orbits are elongated rather circular, but the
enhancement of dark matter density in the luminous parts of galaxies
is robust.  The number of galaxies studied at this level of detail
remains small, but all available analytical and numerical studies show
that enhanced density of dark matter halos by dissipative baryons is a
generic result, which is not sensitive to the assumption of slow,
smooth growth of the central galaxy component \citep{jesseit_etal02,
sellwood_mcgaugh05, choi_etal06, maccio_etal06, weinberg_etal06}.  We
are not aware of any hydrodynamic simulations of galaxy formation that
do not show this effect.

The inclusion of AC amplifies the effects of disk gravity, making it
difficult to reproduce the observed {\it lack} of correlation between
disk size and rotation speed \citep{courteau_rix99}.  By boosting disk
rotation speeds relative to halo virial velocities, AC also makes it
difficult to reconcile the observed Tully-Fisher relation with the
observed galaxy luminosity function, given the halo population of
typical CDM models; this reconciliation is a long-standing challenge
to semi-analytic models of galaxy formation
\citep[e.g.,][]{kauffmann_etal93, cole_etal94} because the conflict
emerges from a relatively simple, halo-counting argument.  These
observational challenges make it tempting to simply omit AC when
creating models of the galaxy population
\citep[e.g.,][]{somerville_primack99, dutton_etal05, dutton_etal06}.
Given the numerical simulation results cited above, however, we regard
omitting AC as a radical departure from a well established element of
galaxy formation physics.  It is a departure that must be considered
if the observations require it, but one should consider comparably
radical changes (e.g., to cosmological parameters or the stellar
initial mass function) on an equal basis.  Throughout this paper, we
compare models with AC (computed using the modified model of
\citealt{gnedin_etal04}) and without AC, but we regard the former as
``standard galaxy formation physics'' and the latter as a speculative
scenario.

\section{Observational Results}

We analyze the sample of 81 disk-dominated galaxies presented by
\citet{pizagno_etal05}.  The galaxies were selected from the main
spectroscopic galaxy sample \citep{strauss_etal02} of the SDSS
redshift survey and cover a range of absolute magnitudes $-18.5 >
M_{r} > -23.2$, with an approximately flat distribution.  The surface
brightness profiles of these galaxies in the $i$-band were fitted by a
sum of an inclined exponential disk and a \citet{sersic68} bulge, and
all the galaxies in our sample were chosen to have the disk-to-total
luminosity ratio $\ge 0.9$.  This morphological cut, more stringent
than in most TF studies, allows us to define scale length and velocity
measures that are insensitive to ambiguities of disk-bulge
decomposition.  It also makes clear the population we are modeling:
nearly bulgeless galaxies.  Rotation curves for these galaxies were
obtained using long-slit H$\alpha$ spectroscopy on the Calar Alto
3.5-m telescope and the MDM 2.4-m telescope.

We take the measurements of stellar mass, disk scale length, and
rotation speed from \citet{pizagno_etal05}.  As our measure of
rotation speed, we adopt the circular velocity at 2.2 disk scale
lengths, $V_{2.2}$, which according to \citet{courteau97} produces the
tightest TF relation.\footnotemark
\footnotetext{Our data contain small corrections to the observational
uncertainty of $V_{2.2}$ compared to \citet{pizagno_etal05}, as
described by Pizagno et al. (2006, in prep).}
Stellar masses are estimated from the SDSS $i$-band Petrosian
luminosities and the SDSS $g-r$ model colors, both corrected for
internal extinction.  We use the models of \citet{bell_etal03} to
compute stellar mass-to-light ratios as a function of $g-r$ color,
except that we convert from their ``diet Salpeter'' stellar initial
mass function (IMF), which is chosen so that stellar disks have the
maximum mass allowed by rotation curve constraints, to a
\citet{kroupa01} IMF, which better represents direct observational
estimates of the IMF.  This conversion lowers the stellar
mass-to-light ratios by 0.15 dex.  Thus, throughout most of this
paper, a galaxy's ``stellar mass'' is $M_* = M_{\ast}/L_i \times L_i$,
where $L_i$ is the inclination-corrected Petrosian $i$-band luminosity
and
\begin{equation}
  \log{M_{\ast}/L_i} = -0.222 + 0.864 \, (g-r) - 0.15.
  \label{eq:mtol}
\end{equation}
We will also consider a model with a ``light'' IMF, in which the
stellar mass at fixed $L_i$ and $g-r$ is 0.15 dex lower (see
\S\ref{sec:mdscatter}).  Galaxy distances, which affect the stellar
mass and scale length values, are computed from redshifts assuming
$H_0 = 70$ km s$^{-1}$ Mpc$^{-1}$.  The observational uncertainties in
$M_*$ and $R_d$ incorporate a $1\sigma$ distance uncertainty
corresponding to 300 km s$^{-1}$, the typical amplitude of small scale
deviations from Hubble flow \citep{strauss_willick95}.  As emphasized
by Pizagno et al. (2006) and evident in the plots below, an important
feature of this sample relative to many previous TF samples is that
observational uncertainties are small compared to the intrinsic
scatter, because axis ratios are precisely measured and because the
minimum redshift $cz = 5000$ km s$^{-1}$ ensures that fractional
distance uncertainties are small.  Throughout this paper, we use the
notation {\it log} for a base-10 logarithm and {\it ln} for a natural
logarithm.  We use the units of solar mass for $M_*$, kpc for $R_d$,
and km s$^{-1}$ for $V_{2.2}$.

\begin{figure}[t]
\vspace{-0.5cm}
\centerline{\epsfysize3.5truein \epsffile{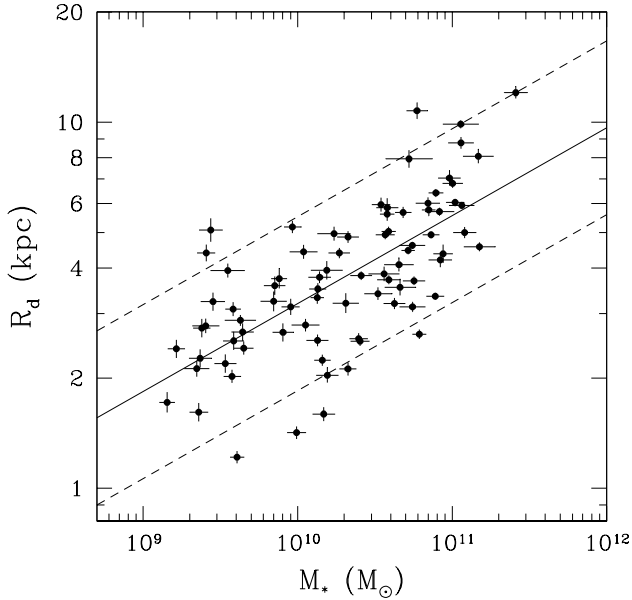}}
\vspace{-0.3cm}
\caption{Stellar mass -- disk scale length relation for the observed
  sample.  Dashed lines are displaced vertically from the mean
  relation ({\it solid line}) by $\pm 1.65\, \sigma$, where $\sigma$
  is the intrinsic scatter.  For a Gaussian distribution around the
  mean relation with standard deviation $\sigma$, dashed lines would
  enclose 90\% of the data points, in the absence of observational
  errors (which modestly increase the dispersion relative to the
  intrinsic scatter).}
  \label{fig:mr_obs}
\end{figure}

\begin{figure*}[t]
\vspace{-0.5cm}
\centerline{\epsfysize3.5truein \epsffile{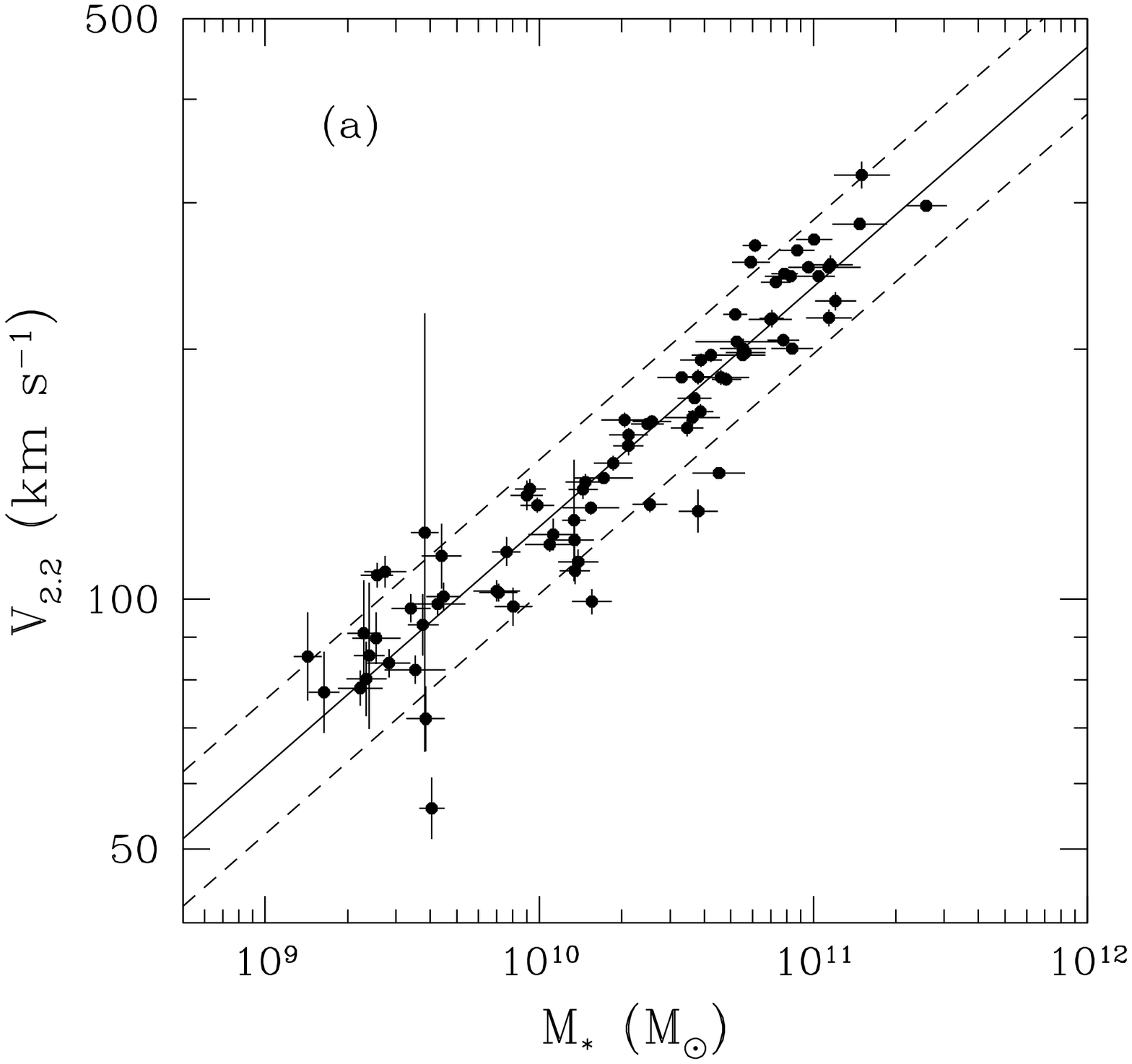}
            \epsfysize3.5truein \epsffile{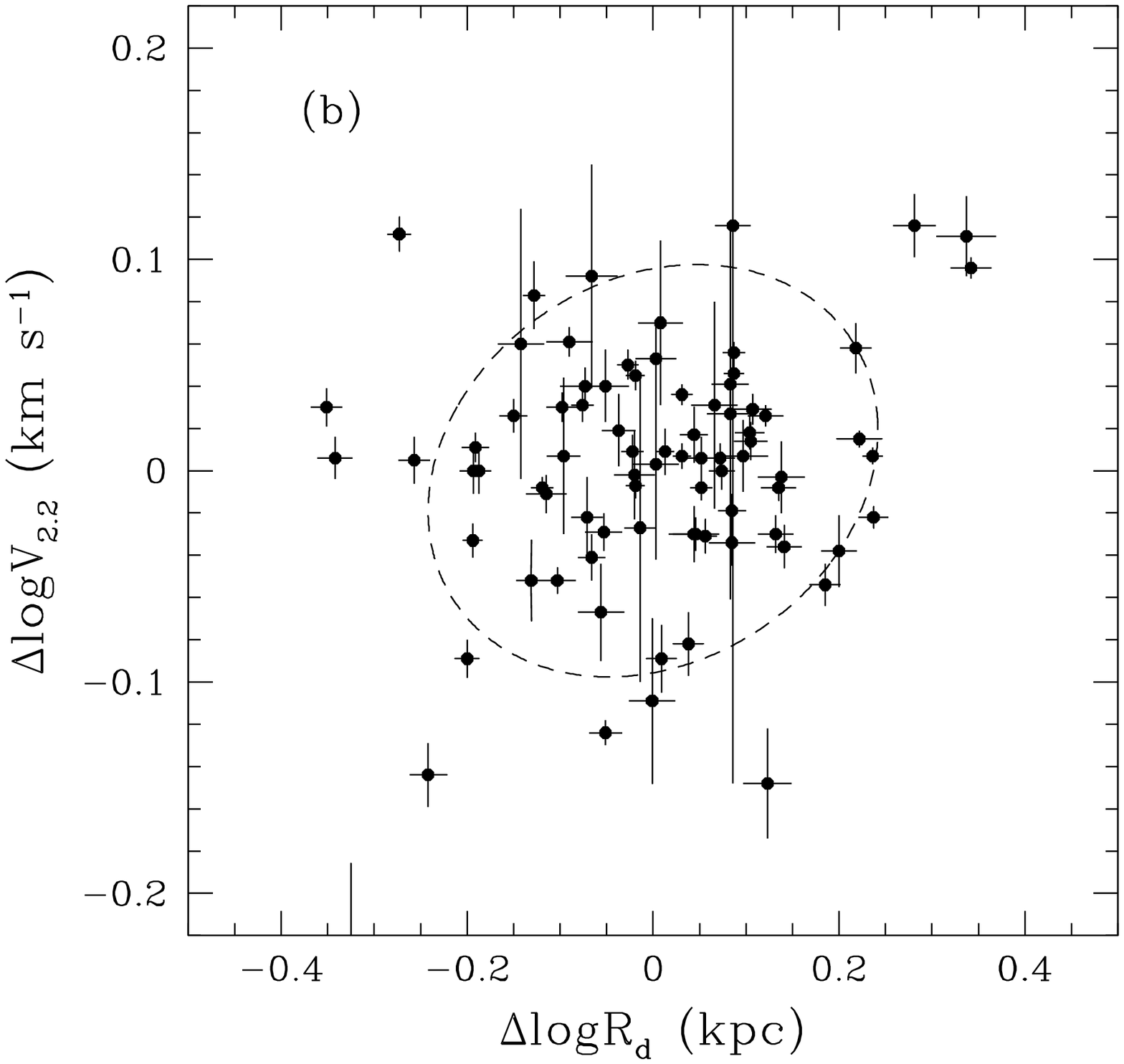}}
\vspace{-0.3cm}
\caption{{\it Left panel:} Observed TF relation in the $i$ band with
  the \protect\citet{bell_etal03} mass-to-light ratios modified to a
  Kroupa IMF.  Dashed lines are displaced vertically from the mean
  relation by $\pm 1.65\, \sigma$, where $\sigma$ is the intrinsic
  scatter.  For a Gaussian distribution with standard deviation
  $\sigma$, the two dashed lines would enclose 90\% of data points, in
  the absence of observational errors.  {\it Right panel:} Residuals
  of the TF relation, $\Delta V_{2.2}$, vs. residuals of the mass-size
  relation, $\Delta R_d$.  The dashed ellipse is oriented in the
  direction of the residual correlation, with its major and minor axes
  equal to 1.65 times the rms deviation in the parallel and
  perpendicular directions, respectively.  This ellipse would enclose
  81\% of data points if the distribution was Gaussian.}
  \label{fig:tf_obs}
\end{figure*}

Figure~\ref{fig:mr_obs} shows the distribution of sample galaxies in
the plane of $R_d$ and $M_*$.  Figure \ref{fig:tf_obs}a shows the
$V_{2.2} - M_*$ relation.  We calculate the parameters of a linear fit
\begin{equation}
  \log{V_{2.2}} = a (\log{M_*} - \log{M_{*0}}) + b + N(0,\sigma),
  \label{eq:tf}
\end{equation}
using the maximum likelihood method described in Appendix.  The last
term in equation (\ref{eq:tf}) allows for an intrinsic Gaussian
scatter of the relationship with standard deviation $\sigma$.  The
offset, $\log{M_{*0}} = 10.5$, is chosen to minimize the error of the
intercept, $\Delta b$, and to eliminate correlation between $\Delta a$
and $\Delta b$.  Table \ref{tab:tf} lists the best-fit parameters: $a
= 0.291 \pm 0.013$, $b = 2.231 \pm 0.006$, $\sigma = 0.049 \pm 0.007$.
The slope and intrinsic scatter are identical within the errors to
those given in \citet{pizagno_etal05}, who used a different
implementation of the same method.  Our intercept is different because
of the slightly different choice for the offset $\log{M_{*0}}$.  We
have verified that our method reproduces the values $b$ and $\Delta b$
of \citet{pizagno_etal05} given their choice of $\log{M_{*0}}$.

As Figure~\ref{fig:mr_obs} shows, the disk scale length correlates
strongly with the stellar mass.  The average relation can be
determined by the same ML method:
\begin{equation}
  \log{R_d} = 0.62 + 0.24 (\log{M_*} - 10.5).
  \label{eq:r-m}
\end{equation}
We then remove the mean mass dependence of the velocities and sizes,
and look at the distribution of residuals, $\Delta V_{2.2}$
vs. $\Delta R_d$.  Figure \ref{fig:tf_obs}b shows that this
distribution is essentially a scatter plot, with at most a slight
positive correlation (Spearman's linear correlation coefficient $r =
0.23 \pm 0.14$).  \citet{courteau_rix99} emphasized that the lack of
correlation between TF residuals and disk scale length argues against
the ``maximal disk'' hypothesis, in which the stellar disk provides a
large fraction of the rotational support at $2.2R_d$; with $V^2 \sim
GM_*/R$, more compact disks would rotate faster at fixed $M_*$,
creating a strong negative correlation between $\Delta V_{2.2}$ and
$\Delta R_d$.  The fact that the residual data reveal only a weak
positive correlation, rather than a substantial negative correlation,
will play a key role in our conclusions below about the baryon
fractions of disk galaxies.

We have also investigated whether a fundamental plane-type fit as a
function of two variables, $M_*$ and $R_d$, would lead to a smaller
intrinsic scatter of $V_{2.2}$.  We use our maximum likelihood method
to fit a linear relation of the form
\begin{equation}
  \log{V_{2.2}} = a_1 (\log{M_*} - \log{M_{*0}}) + 
                  a_2 (\log{R_d} - \log{R_{d0}}) + b + N(0,\sigma).
  \label{eq:fp}
\end{equation}
The error estimates are uncorrelated for $\log{M_{*0}} = 10.5$ and
$\log{R_{d0}} = 0.65$.  The best-fit parameters are given in Table
\ref{tab:fp}: $a_1 = 0.271 \pm 0.017$, $a_2 = 0.085 \pm 0.054$, $b =
2.233 \pm 0.006$.  However, the estimated intrinsic scatter of this
relation is essentially the same as the intrinsic scatter of the TF
relation above (0.048 vs. 0.049 dex), which supports the notion that
the TF relation is already a nearly edge-on view of the fundamental
plane.  The second, radius-related exponent $a_2$ is statistically
consistent with zero, and it is much smaller than the first,
mass-related exponent.  The relative unimportance of the disk size is
another manifestation of the lack of residual correlation in Figure
\ref{fig:tf_obs}b, and the sign of the $a_2$ coefficient is again
opposite to the expectation for self-gravitating disks.  As a
consistency check, if we substitute the mean mass-size relation
(\ref{eq:r-m}) into equation (\ref{eq:fp}), we recover the original TF
relation (eq. [\ref{eq:tf}]).

\section{Models}
  \label{sec:models}

In order to interpret these data, we construct disk galaxy models
within the framework of a hierarchical $\Lambda$CDM cosmology.  We
assume that each galaxy resides within a dark matter halo whose virial
radius, $R_{\rm vir}$, encloses a mean overdensity $\Delta = 101$
relative to the critical matter density, appropriate for a flat
cosmological model with $\Omega_{\rm m} \approx 0.3$
\citep{bryan_norman98}.
%Even though the latest WMAP parameters prefer a slightly lower value
%(93), it affects the virial radii and predicted spin parameters by
%less than 3\%.
We assume that the mass profiles of these halos are given by an NFW
model \citep{navarro_etal97}, which is parametrized by the virial
mass, $M$, and concentration parameter, $c$.  Halos found in
dissipationless $\Lambda$CDM simulations have concentrations whose
mean $\bar{c}$ correlates with halo mass and whose distribution has a
Gaussian scatter in $\log{c}$:
\begin{equation}
  P(c) \, d\log{c} = {1 \over \sqrt{2\pi} \sigma_c} \,
    \exp{\left(-{[\log{c} - \log{\bar{c}(M)}]^2 \over 2 \sigma_c^2}\right)} \,
    d\log{c}.
  \label{eq:c}
\end{equation}
\citet{bullock_etal01} find an empirical fit $\bar{c}(M) \approx 12.8
\, (M/10^{12}\, h^{-1} \Msun)^{-0.13}$ for $N$-body simulations with
$\Omega_{\rm m} = 1-\Omega_{\rm \Lambda} = 0.3$ and $\sigma_8=1.0$.
We adopt this $\bar{c}(M)$ relation as our standard, but in
\S\ref{sec:lf} we also consider the lower concentrations predicted by
the \citet{bullock_etal01} analytical model in the case of lower
$\sigma_8$.  The scatter is $\sigma_c = 0.14$ dex for all halos, and
may be smaller for those halos that have not had a recent major merger
\citep{wechsler_etal02}.  As discussed in the introduction, our
standard models include the effect of contraction of dark matter halos
in response to condensation of baryons in their centers
\citep{barnes_white84, blumenthal_etal86, ryden_gunn87}, computed
using the modified adiabatic contraction model of
\citet{gnedin_etal04}.  We also explore models in which adiabatic
contraction does not occur, although we consider these less physically
reasonable.

We treat the stellar component of each galaxy as an exponential disk
with the measured mass $M_*$ and scale length $R_d$; since disk-bulge
decomposition yields a bulge-to-total ratio $\leq 0.1$ for all
galaxies in our sample, this assumption is appropriate.  We do not
have gas masses available for our sample, and therefore we omit the
gas contribution to the disk potential and denote the ratio of the
disk {\it stellar} mass to the halo virial mass by $m_d$.  In many
theoretical papers, $m_d$ represents the fraction of disk {\it baryon}
mass (stars plus cold gas).  Since gas fractions are typically small
in the luminosity range that we consider, the difference between the
two definitions should only be $15\%$ on average (based on the
\citealt{kannappan04} relation for $u-r$ galaxy colors).

The choice of $m_d$ and $c$ completely specifies the model of an
individual observed galaxy, given $M_*$ and $R_d$.  The angular
momentum parameter $\lambda_d$ is not an input, but can instead be
derived from the total disk angular momentum:
\begin{equation}
  J_d \equiv \lambda_d \, \sqrt{2} M_* R_{\rm vir} V_{\rm vir}
      = \int V_c(R) R \, 2\pi\Sigma(R) R dR,
  \label{eq:lamd}
\end{equation}
where $\Sigma(R)$ is the surface mass density of the disk at azimuthal
radius $R$, and $V_c(R)$ is the rotation velocity in the plane of the
disk, which includes the contributions of the flat disk and the
spherical halo.  Note that we use the {\it model} $V_c(R)$ in
computing $\lambda_d$ from equation (\ref{eq:lamd}), and a surface
density profile $\Sigma(R)$ that is exponential with the observed
$M_*$ and $R_d$.  The virial velocity is defined as $V_{\rm vir}^2 =
GM/ R_{\rm vir}$, where $R_{\rm vir}$ is the virial radius.  The
angular momentum of the halo can be similarly expressed using the halo
spin parameter $\lambda$:
\begin{equation}
  J_h \equiv \lambda \, \sqrt{2} M R_{\rm vir} V_{\rm vir}.
  \label{eq:lamh}
\end{equation}

In our framework, a model of the disk galaxy population is defined by
the probability distribution $P(m_d | M_*,R_d)$, and our goal is to
investigate what forms of this distribution yield an acceptable fit to
the observed distribution of galaxies in the ($M_*, R_d, V_{2.2}$)
space.  For each individual galaxy of the \cite{pizagno_etal05}
sample, we proceed as follows.  (1) We draw a disk fraction, $m_d$,
from the assumed distribution $P(m_d | M_*,R_d)$ and calculate the
halo virial mass, $M = M_*/m_d$, using the observed stellar mass
$M_*$.  (2) We draw an NFW concentration parameter, $c$, from the
theoretical distribution $P(c)$, given by equation (\ref{eq:c}).  In
some models we set the scatter of concentrations to $\sigma_c = 0$.
(3) We calculate the adiabatically-contracted halo mass profile using
the modified model of AC with the known $M_*$, $M$, $c$, and $R_d$.
We also run models omitting the AC effect.  (4) We calculate the
rotation velocity at $2.2 \, R_d$ and add random measurement errors to
$\log{M_*}$, $\log{R_d}$, and $\log{V_{2.2}}$, drawn from their
respective Gaussian distributions, each with the standard deviation
equal to the corresponding uncertainty for the observed galaxy.  This
step is necessary in order to correctly model the correlation of
residuals.

\begin{figure*}[t]
\vspace{-0.0cm}
\centerline{\epsfysize5.5truein \epsffile{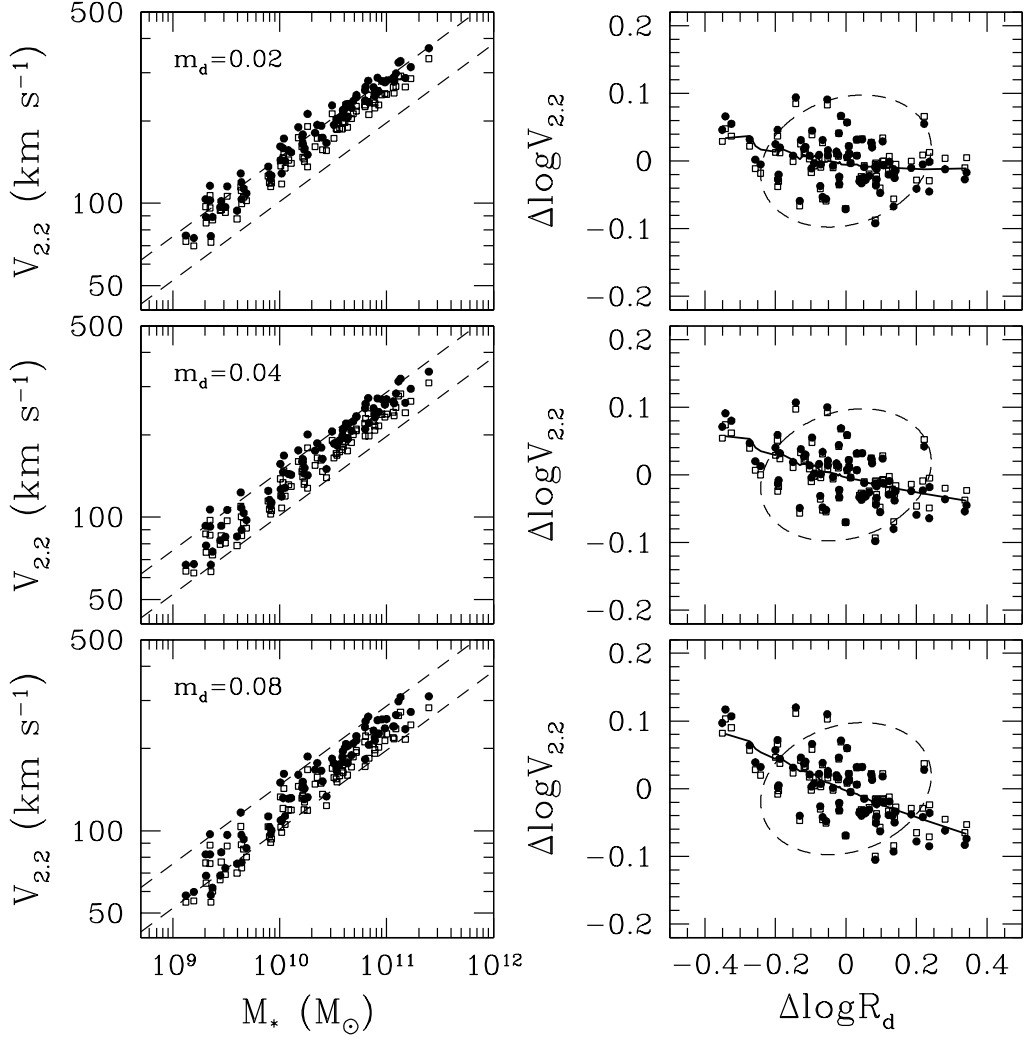}}
\vspace{-0.0cm}
\caption{TF relation and residuals for three representative models
  with constant $m_d$ and no scatter in halo concentrations,
  $\sigma_c=0$.  Filled circles are for models including the effect of
  adiabatic contraction of dark matter, open squares are for models
  without the AC effect.  Dashed lines ({\it left panel}) and contours
  ({\it right panel}) are for the data from
  Fig. \protect\ref{fig:tf_obs}.  Scatter in the right panels is
  almost entirely caused by the measurement errors added to model
  values of $M_*$, $R_d$, and $V_{2.2}$.  The solid lines in the right
  panels show the original model realizations without including the
  measurement errors.  These lines correspond to the case with AC, the
  case without AC is similar.}
  \label{fig:tf_const}
\end{figure*}

\begin{figure*}[t]
\vspace{-4.2cm}
\centerline{\epsfysize5.5truein \epsffile{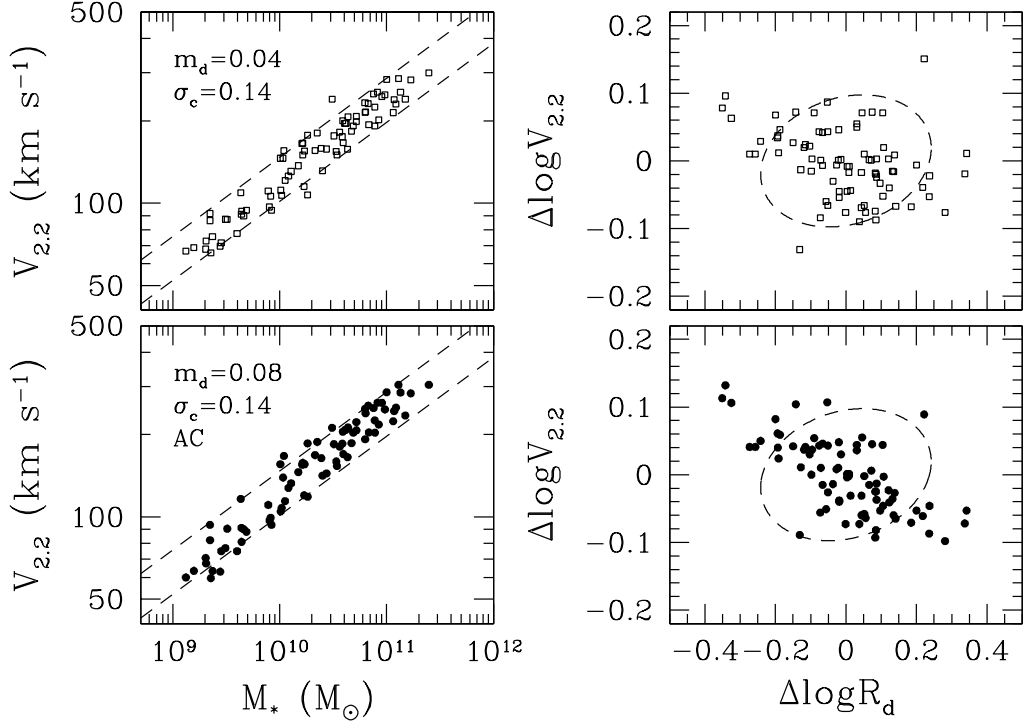}}
\vspace{-0.0cm}
\caption{Best-fit models to the TF relation and residuals with
  constant $m_d$, including the scatter of halo concentrations,
  $\sigma_c=0.14$.  {\it Bottom panel:} models including the effect of
  halo contraction, with $m_d=0.08$; {\it top panels:} without the AC
  effect, with $m_d=0.04$.}
  \label{fig:tf_sigc}
\end{figure*}

% \begin{figure}[t]
% \vspace{-3.5cm}
% \centerline{\epsfysize5.9truein \epsffile{f5.eps}}
% \vspace{-0.2cm}
% \caption{Best-fit models to the TF relation and residuals with
%   constant $m_d$ and scatter of halo concentrations, with added
%   velocity errors to match the observed intrinsic scatter.  {\it
%   Bottom panel:} models including the effect of halo contraction, with
%   $m_d=0.08$; {\it top panels:} without the AC effect, with
%   $m_d=0.04$.}
%   \label{fig:tf_addsig}
% \end{figure}

In constructing our model sample, we implicitly assume that the model
predictions can be calculated with sufficient accuracy and that the
only uncertainty in comparing with the data is due to the measurement
errors.  However, some of our models are probabilistic in nature, as
they contain scatter of halo concentrations and disk fractions.  Given
only 81 galaxies in the observed sample, the predictions of these
models vary from one realization to another.  Even for fixed $m_d$ and
$\bar{c}(M)$, the random errors added to $\log{M_*}$ and
$\log{V_{2.2}}$ lead to small variations in best-fit parameters among
realizations.  In order to suppress such statistical variations and
obtain robust average model predictions, we take a large number of
repetitions of each observed galaxy, $N_{\rm rep}$, and fit a single
TF relation (and FP relation) to the combined sample of $81 \times
N_{\rm rep}$ galaxies.  We have conducted a convergence study, by
increasing $N_{\rm rep}$ until the model fit parameters vary by less
than 10\% of the corresponding observational uncertainty.  We have
found that this condition requires at least 100 realizations of the
sample in some cases, and we adopt $N_{\rm rep} = 100$ for all our
models.

We fit the TF relation to our sample of 8100 model galaxies as in
equation (\ref{eq:tf}), with the same $\log{M_{*0}}$ used for the
observational data.  We compute $\Delta V_{2.2}$ and $\Delta R_d$
residuals relative to this best-fit TF relation and compute the
correlation coefficient as for the observational data.  The best-fit
slopes $a$, zero-points $b$, intrinsic dispersions $\sigma$, and
correlation coefficients $r$ of all our model samples are given in
Table \ref{tab:tf}.  We also fit the FP relation similarly and list
the best-fit parameters in Table \ref{tab:fp}.

For a given model sample, we evaluate the goodness-of-fit to the
observed TF relation based on the $\chi^2$ statistic:
\begin{eqnarray}
  \chi^2 & = & \left({a_{\rm mod}-a_{\rm obs} \over \Delta a_{\rm obs}}\right)^2
         + \left({b_{\rm mod}-b_{\rm obs} \over \Delta b_{\rm obs}}\right)^2
         + \left({\sigma_{\rm mod}-\sigma_{\rm obs} \over \Delta \sigma_{\rm obs}}\right)^2\nonumber\\
         & + & \left({r_{\rm mod}-r_{\rm obs} \over \Delta r_{\rm obs}}\right)^2.
\end{eqnarray}
For the goodness-of-fit of the FP relation, we use
\begin{eqnarray}
  \chi^2 & = & \left({a_{1,\rm mod}-a_{1,\rm obs} \over \Delta a_{1,\rm obs}}\right)^2
         + \left({a_{2,\rm mod}-a_{2,\rm obs} \over \Delta a_{2,\rm obs}}\right)^2
         + \left({b_{\rm mod}-b_{\rm obs} \over \Delta b_{\rm obs}}\right)^2\nonumber\\
         & + & \left({\sigma_{\rm mod}-\sigma_{\rm obs} \over \Delta \sigma_{\rm obs}}\right)^2.
\end{eqnarray}
Note that all of our models reproduce the observed ($M_*, R_d$)
distribution by construction.

\subsection{Constant disk mass fraction}

First, we explore models with a constant disk fraction $m_d$ and no
scatter of halo concentrations, $\sigma_c=0$.  In this case the model
predictions for each galaxy are deterministic, except for the added
measurement errors.  Figure~\ref{fig:tf_const} shows the TF relation
for model galaxies and the distribution of residuals.  The top panels
correspond to a low disk mass fraction, $m_d = 0.02$.  For a given
galaxy stellar mass, low $m_d$ corresponds to a large dark matter halo
and, therefore, a high circular velocity.  The model velocities lie
consistently above the observed mean relation, indicating that the
true average value of $m_d$ is larger than 0.02.  Since the halo
dominates the mass even at small radii, the adiabatic contraction
effect is not very important, and the model realizations with and
without the AC effect are similar to each other.

% Note that although we focus in this paper only on a single value of
% the rotation velocity at $2.2 R_d$, the whole shape of the rotation
% curve would be at odds with observations if it were dominated by the
% halo at all radii.

Since the disk contribution to the rotation curve is small, the disk
size has little impact on the velocity residuals (top right panel of
Fig.~\ref{fig:tf_const}).  Most of the plotted velocity residuals
arise from the random ``measurement'' errors added to the model galaxy
values.  A solid line in that panel shows the original predicted
relation of the residuals, before we add the random errors.  If the
disk completely dominated the rotation curve, at a given stellar mass
we would expect an anti-correlation of the form $\Delta\log{V}_{2.2} =
-0.5 \, \Delta\log{R_d}$.  With a large contribution of the dark halo,
the relation of the residuals is much flatter but is still
approximately monotonic.  The relation plotted by the solid line is
insensitive to the stellar mass -- we have split the sample in bins of
large and small $M_*$ and found approximately the same relation as for
the whole sample.

In statistical terms, the $m_d = 0.02$ models have acceptable TF
slopes but zero-points that are too high, $b = 2.306$ (2.265) with
(without) AC, compared to the observed values of $2.231 \pm 0.006$.
Furthermore, both models have an intrinsic scatter that is too small,
$\sigma = 0.014$ (0.009) compared to the observed $\sigma = 0.049 \pm
0.007$.

The middle and bottom rows of Fig.~\ref{fig:tf_const} show model
realizations for larger disk mass fractions, $m_d = 0.04$ and $m_d =
0.08$, respectively.  Increasing $m_d$ lowers the halo mass, given the
observed $M_*$, and thus lowers the TF zero-point, bringing the models
into better accord with the data.

Within this class of constant--$m_d$, constant--$c$ models, the best
$\chi^2$ fit is achieved for $m_d \approx 0.08$ with AC and $m_d
\approx 0.04$ without; AC boosts $V_{2.2}$ and therefore requires
higher $m_d$.  However, both models are statistically unacceptable,
with total $\chi^2$ of 36 and 46, respectively (for four observables
and one adjustable parameter, hence three degrees of freedom).  The
no--AC model has a slope that is slightly too steep (0.295 vs. $0.291
\pm 0.013$), but its more serious problems are predicting too little
scatter ($\sigma = 0.016$ vs. $0.049 \pm 0.007$) and predicting
anti-correlated residuals (coefficient $r = -0.26$ vs. $+0.23 \pm
0.14$).  Disk gravity plays a much larger role in the AC model,
because of both the higher $m_d$ and the effects of AC itself.  The
slope discrepancy becomes worse ($a = 0.322$) because more massive
disks are higher surface density on average and produce stronger
boosts in $V_{2.2}$.  The intrinsic scatter is closer to the observed
value ($\sigma = 0.034$), but this scatter is driven entirely by
variations in disk size, so the anti-correlation of residuals is worse
($r = -0.57$).

The conclusions of the FP analysis are similar.  The mass coefficient
$a_1$ increases and the intercept $b$ decreases with increasing $m_d$.
The radius coefficient $a_2$ becomes more negative, further away from
the observed value, the signature of an overly dominant disk.  The
intrinsic scatter of the FP relation does not vary noticeably with
$m_d$, but it is much lower than observed, $\sigma \approx 0.008$
vs. $0.048 \pm 0.006$ (see Table \ref{tab:fp}).

It is easy to see why models with fixed values of $m_d$ and $c$ cannot
reproduce the observed intrinsic scatter and uncorrelated residuals,
regardless of the value of $m_d$ and the presence or absence of AC.
In these models, the pre-contraction halo profile is determined by
$M_*$, and the variation in $R_d$ is therefore the only available
source of intrinsic scatter of $V_{2.2}$ at fixed $M_*$.  Thus, if
there is significant intrinsic scatter, there {\it must} be strong TF
residual anti-correlations.  The FP analysis removes the scatter
associated with $R_d$, making $\sigma$ independent of $m_d$ (and much
too small).

Of course, dark matter halos in cosmological simulations display a
range of concentrations at fixed virial mass, characterized by the
log-normal distribution of equation (\ref{eq:c}) with $\sigma_c =
0.14$ dex.  We add this variation to our second set of models, plotted
in Figure~\ref{fig:tf_sigc}.  The scatter of halo profiles
significantly increases the intrinsic scatter of the model TF relation
(see Table \ref{tab:tf}), especially for low $m_d$ models, and brings
it within striking range of the observed value.  Since the
concentration variations are uncorrelated with $R_d$, they also reduce
the anti-correlation of TF residuals.  Both effects improve $\chi^2$,
but they still do not produce acceptable fits to the data, for either
the TF or FP relations.  The best models have total $\chi^2$ of 40
with AC and 13 without (Table \ref{tab:tf}, TF) and larger values for
FP (Table \ref{tab:fp}), again for three degrees of freedom.

Note that in the \citet{mo_etal98} formalism, the concentration
parameter encodes the effect of halo formation history on halo
profile, essentially the effect described as {\it formation redshift}
in earlier work \citep[e.g.,][]{cole_kaiser89, eisenstein_loeb96}.
While these earlier papers highlighted the problem of over-predicting
the scatter of the TF relation, because of variations in formation
redshift, the range of concentrations found in cosmological {\it
N}-body simulations is actually insufficient in itself to explain the
observed scatter.  Furthermore, since the tail of low concentrations
comes from halos with active late-time merger histories, we consider
it plausible that the concentration parameters of halos that host
disk-dominated galaxies have smaller scatter than $\sigma_c = 0.14$,
perhaps much smaller.

Suppose that we have systematically underestimated the
``observational'' errors in $V_{2.2}$ and $M_*$, perhaps because of
non-circular motions, disk ellipticities that affect inclination
corrections \citep{franx_dezeeuw92}, or variations in stellar
mass-to-light ratio at fixed color (present at the 0.1 dex level in
\citealt{bell_etal03}).  We put ``observational'' in quotes because
these could equally well be considered ``physical'' sources of
scatter, but they are in any event not included in the models above.
We consider an upper limit on these effects by adding random velocity
errors to the models until the intrinsic scatter of the model
relationship matches that of the observed TF relation.  In practice,
we add log-normally distributed errors in $V_{2.2}$ with the variance
equal to $\sigma_{\rm obs}^2 - \sigma_{\rm mod}^2$.  The TF intrinsic
scatter matches the observed value by construction, and the FP scatter
is close to that observed.  However, the TF residual anti-correlations
remain much too strong, and the $a_2$ slopes of the FP fits remain
negative and large (Tables \ref{tab:tf} and \ref{tab:fp}).
Realizations of these models (not shown) are visually similar to
Figure \ref{fig:tf_sigc}.

% However, Figure~\ref{fig:tf_addsig} shows that adding
% this random scatter has very little effect on the residual correlation
% and may even make it worse.  The $\chi^2$ of the model fits are also
% not improved, because the slope $a$ is systematically larger than
% observed.

We conclude that models with a single value of $m_d$ cannot reproduce
the observed TF relation and TF residuals of our disk-dominated galaxy
sample, and that much of the intrinsic scatter of the TF relation must
arise from scatter in the ratio of stellar disk mass to halo mass.
\citet{pizagno_etal05} reached the same conclusion based on a more
qualitative comparison of models to the data.  We turn to models with
$m_d$--scatter in the next section, but first we examine the
$\lambda_d$ distributions inferred for our constant $m_d$ models.

\begin{figure}[t]
\vspace{-0.5cm}
\centerline{\epsfysize3.5truein \epsffile{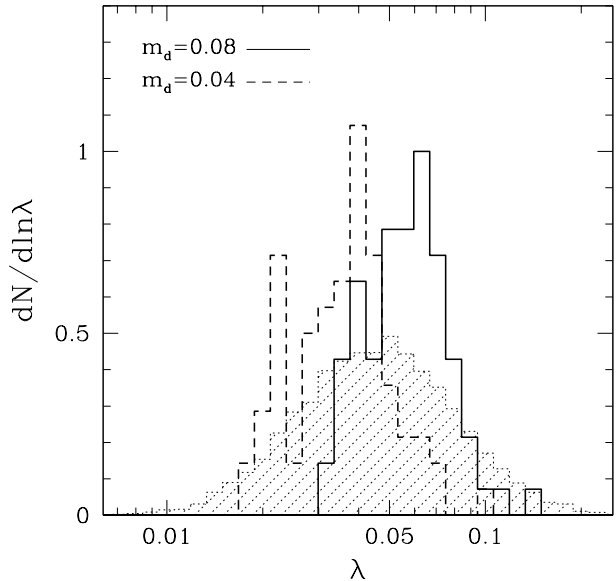}}
\vspace{-0.5cm}
\caption{Distribution of disk angular momentum parameter $\lambda_d$
  required by our models to reproduce the observed disk sizes for the
  models with constant $m_d$, including scatter of halo
  concentrations, $\sigma_c = 0.14$.  Solid histogram is for $m_d =
  0.08$ with AC, dashed histogram is for $m_d = 0.04$ without AC.  For
  comparison, dotted hatched histogram shows the $P(\lambda)$
  distribution of dark matter halos in dissipationless cosmological
  simulations, parametrized by a log-normal function (see
  eq. [\ref{eq:lambda}] in the text).  Normalization is such that the
  peak of the first distribution is unity.}
  \label{fig:lam}
\end{figure}

\begin{figure*}[t]
\vspace{-4.2cm}
\centerline{\epsfysize5.5truein \epsffile{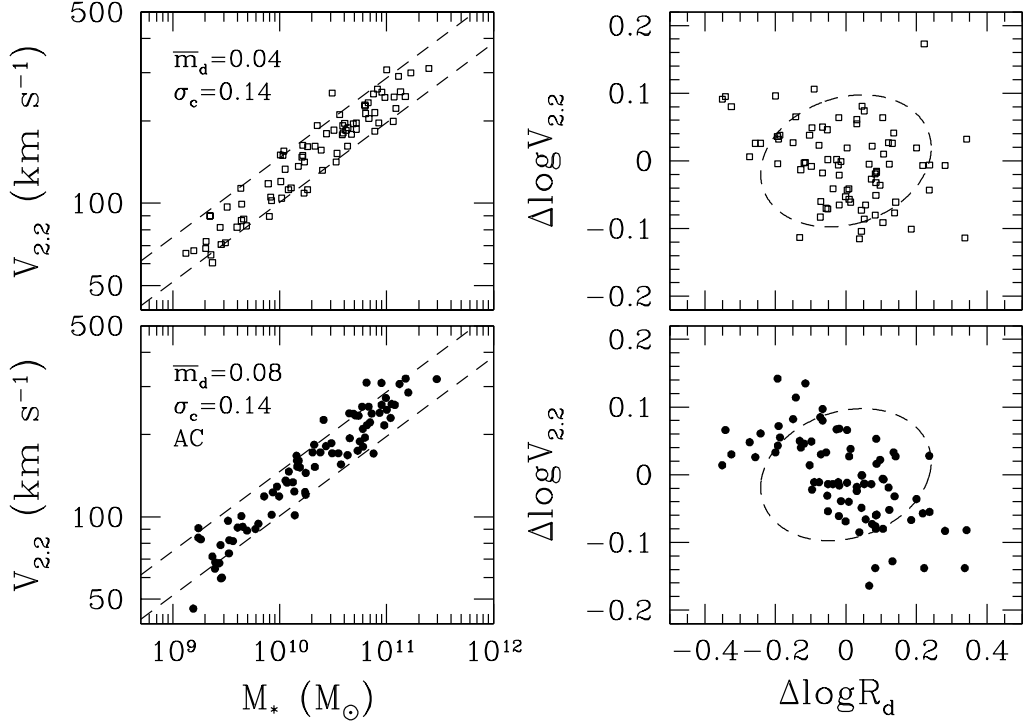}}
\vspace{-0.0cm}
\caption{Best-fit models to the TF relation and residuals with the
  scatter of disk mass fractions, $\sigma_m = 0.22$ ({\it top panel},
  no AC) and $\sigma_m = 0.27$ ({\it bottom panel}, with AC), and
  scatter of halo concentrations, $\sigma_c=0.14$.}
  \label{fig:tf_sigm}
\end{figure*}

\begin{figure*}[t]
\vspace{-4.2cm}
\centerline{\epsfysize5.5truein \epsffile{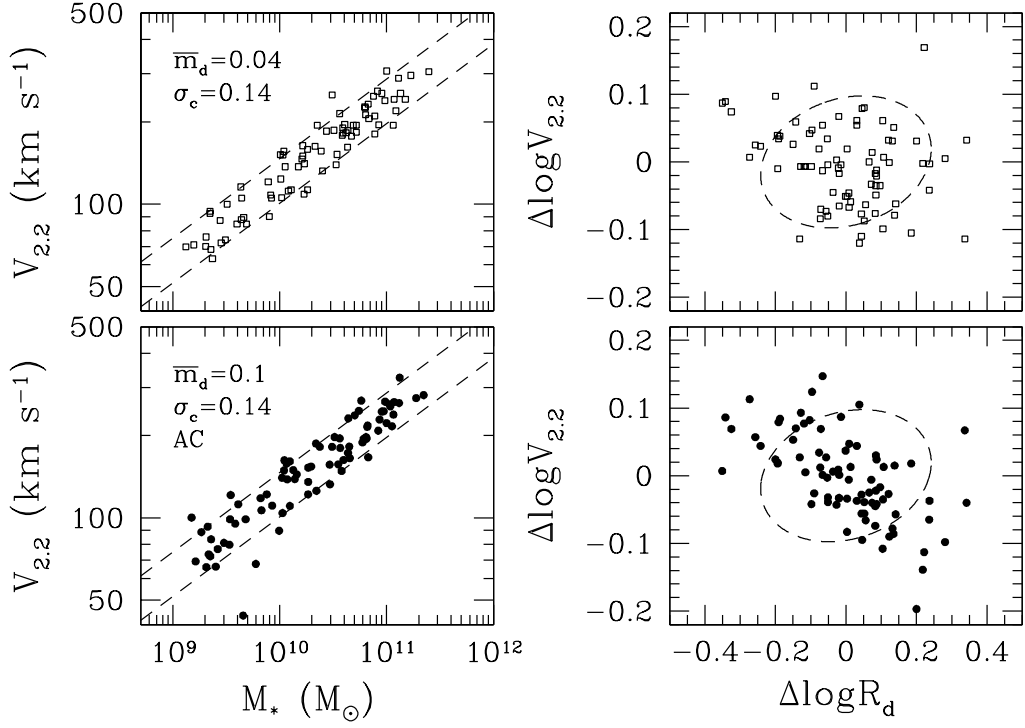}}
\vspace{-0.0cm}
\caption{Best-fit models to the TF relation and residuals with the
  mean disk mass fraction increasing with stellar mass, $\bar{m}_d
  \propto M_*^{p_1}$, including the scatter $\sigma_m$ and
  $\sigma_c=0.14$.  {\it Top panel:} $p_1=0.1$, $\sigma_m=0.25$, no
  AC; {\it bottom panel:} $p_1=0.3$, $\sigma_m=0.21$, with AC.}
  \label{fig:tf_ms}
\end{figure*}

\begin{figure*}[t]
\vspace{-4.2cm}
\centerline{\epsfysize5.5truein \epsffile{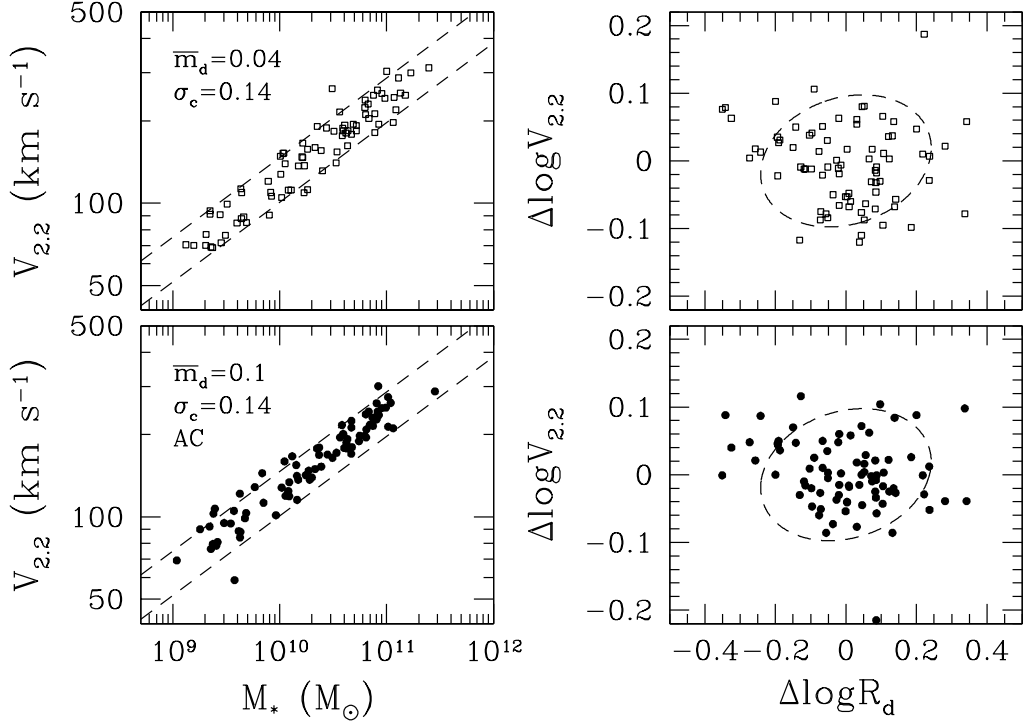}}
\vspace{-0.0cm}
\caption{Best-fit models to the TF relation and residuals with the
  mean disk mass fraction increasing with stellar surface density,
  $\bar{m}_d \propto \Sigma_*^{p_2}$, including the scatter $\sigma_m$
  and $\sigma_c=0.14$.  {\it Top panel:} $p_2=0.2$, $\sigma_m=0.25$,
  no AC; {\it bottom panel:} $p_2=0.65$, $\sigma_m=0.2$, with AC.}
  \label{fig:tf_sden}
\end{figure*}

\begin{figure*}[t]
\vspace{0.2cm}
\centerline{\epsfysize5.5truein \epsffile{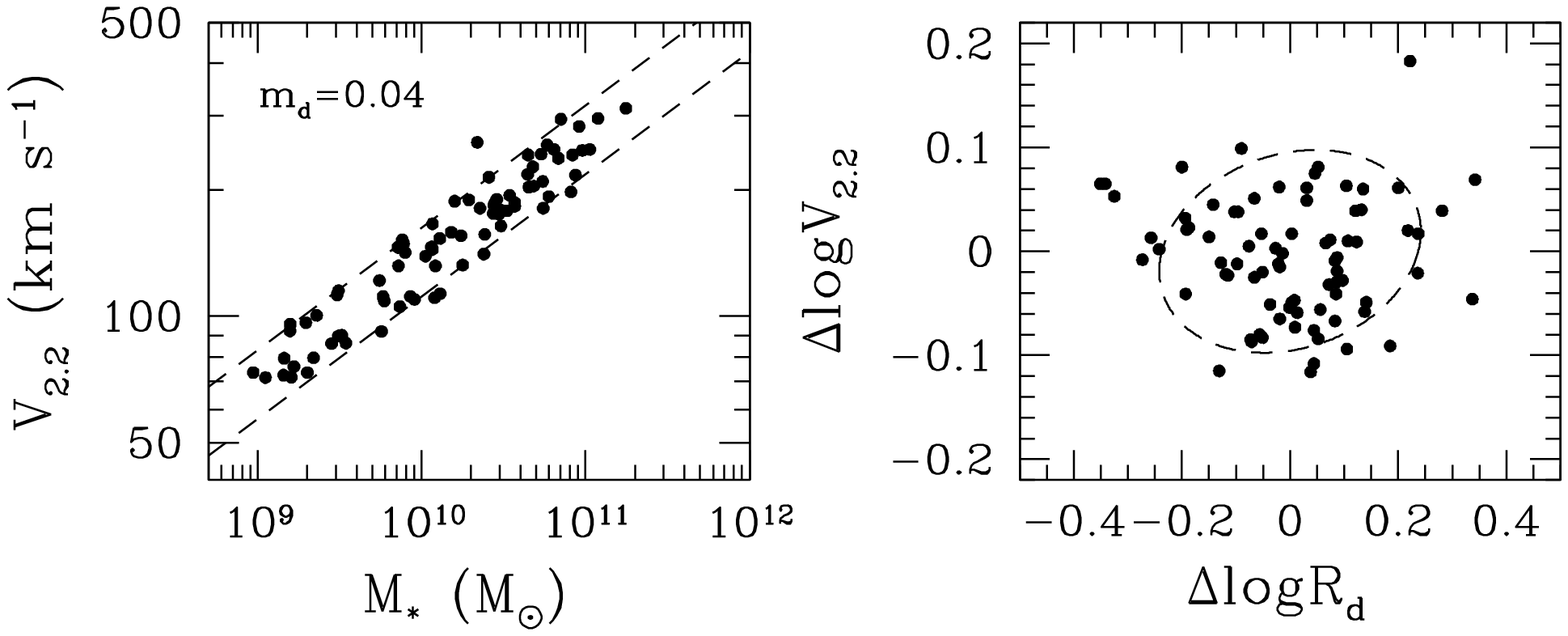}}
\vspace{-8.3cm}
\caption{Best-fit models to the TF relation and residuals for the case
  of light IMF, with the mean disk mass fraction increasing with
  stellar surface density, $\bar{m}_d \propto \Sigma_*^{0.4}$,
  including the scatter $\sigma_m=0.25$ and $\sigma_c=0.14$, and AC.
  The dashed lines representing the observations in the left panel are
  shifted by $-0.15$ dex in $M_*$ to account for the changed IMF.}
  \label{fig:tf_imf1}
\end{figure*}

Figure \ref{fig:lam} shows the $P(\lambda_d)$ distributions for the
two models illustrated in Figure \ref{fig:tf_sigc}, which include
concentration scatter.  In the $m_d = 0.08$, AC model, galaxies reside
in less massive halos with smaller virial radii, so larger $\lambda_d$
values are required to produce their observed scale lengths.  If the
models exactly reproduced each galaxy's observed rotation curve
$V_c(R)$, then equation (\ref{eq:lamd}) implies $\lambda_d \propto
R_{\rm vir}^{-1} V_{\rm vir}^{-1} \propto m_d^{2/3}$ on a
galaxy-by-galaxy basis, independent of AC except for its impact on the
best-fitting $m_d$.  Since the two models approximately reproduce the
observed TF zero-point and slope, they obey this scaling approximately
but not exactly.  The modes of the inferred $\lambda_d$ distributions
are $\bar{\lambda}_d = 0.056$ (AC, $m_d = 0.08$) and $\bar{\lambda}_d
= 0.036$ (no AC, $m_d = 0.04$), and the standard deviations of
$\ln{\lambda_d}$ are $\sigma_{\lambda d} = 0.30$ and $\sigma_{\lambda
d} = 0.36$, respectively.  (In agreement with established notation, we
quote here the standard deviation of the natural logarithm of
$\lambda$; all the other dispersions in this paper are for base-10
logarithms.)

The shaded histogram in Figure~\ref{fig:lam} shows the theoretically
predicted distribution of dark matter {\it halo} spin parameters
(eq. [\ref{eq:lamh}]) derived from cosmological $N$--body simulations
\citep[e.g.,][]{vitvitska_etal02},
\begin{equation}
  P(\lambda) d\lambda = {1 \over \sqrt{2\pi} \sigma_\lambda}
     \exp{\left(-{[\ln{\lambda} - \ln{\bar{\lambda}}]^2 \over 2\sigma_\lambda^2}\right)}
     {d\lambda \over \lambda},
  \label{eq:lambda}
\end{equation}
with $\bar{\lambda} = 0.045$, $\sigma_\lambda = 0.56$.  According to
tidal torque theory \citep[e.g.,][]{hoyle49, peebles69,
efstathiou_jones80}, baryons should acquire the same specific angular
momentum as dark matter at the time of halo turnaround (when most of
the angular momentum is imparted).  However, as noted in the
introduction, the $\lambda_d$ distribution can depart from the
$\lambda$ distribution because disks acquire a biased subset of the
halo baryons, because disk galaxies form in a biased subset of dark
matter halos, because angular momentum is transferred between the dark
matter and the baryons, or perhaps because assembly along a filament
brings baryons from beyond the virial radius.  Figure~\ref{fig:lam}
suggests that one or more of these effects must play an important role
in determining disk angular momenta.  In both of our models, the
inferred $\lambda_d$ distributions are narrower than the theoretical
$\lambda$ distribution.  Interestingly, \citet{wechsler_etal02} and
\citet{donghia_burkert04} find in their cosmological simulations that
halos that have not experienced a recent major merger, and therefore
are likely to host disk-dominated galaxies, have a narrower
distribution $P(\lambda)$ than all halos.  The absence of a
low--$\lambda_d$ tail can plausibly be explained by low-spin halos
producing galaxies with substantial bulges.  For the AC model with
$m_d = 0.08$, the inferred $\lambda_d$ values are systematically
higher than the $N$-body $\lambda$ values.  Since angular momentum
transfer seems more likely to reduce $\lambda_d$ below $\lambda$
\citep{navarro_steinmetz97}, this difference is more plausibly
explained by the other effects mentioned above.

\subsection{Log-normal distributions of disk mass fractions}
  \label{sec:mdscatter}

In the \citet{mo_etal98} formalism of disk galaxy modeling, the
sources of scatter in the TF relation are halo concentration, disk
size, and disk mass fraction.  We have shown in the previous section
that the first two of these predict too little scatter and too much
residual correlation, compared with the data.  Even though boosting
the ``measurement'' errors of model velocities, beyond the level
expected in the data, can in principle explain the observed scatter,
it does not remove the residual correlation.  In this section, we
consider models with a log-normal scatter $\sigma_m$,
\begin{equation}
  P(m_d) \, d\log{m_d} = {1 \over \sqrt{2\pi} \sigma_m} \,
    \exp{\left(-{[\log{m_d} - \log{\bar{m}_d}]^2 \over 2 \sigma_m^2}\right)} \,
    d\log{m_d},
  \label{eq:md}
\end{equation}
and allow $\bar{m}_d$ to scale with stellar mass or stellar surface
density.  In all cases with $m_d$ scatter, we truncate $P(m_d)$ above
the adopted maximum value of 0.15 (close to the universal baryon
fraction) and below the minimum value of 0.001.

We first fix $\bar{m}_d$ at the previous best-fit values for the cases
with and without AC, and increase the dispersion $\sigma_m$ until we
match the observed intrinsic scatter of the TF relation.  This match
requires $\sigma_m \approx 0.2-0.3$.  Figure~\ref{fig:tf_sigm} shows
that even in this case the residual anti-correlation persists,
especially in the realization with AC.  Similarly, in the FP analysis
the slope $a_2$ remains consistently negative, instead of positive as
required by the data.

In addition, the AC model has a significantly discrepant TF slope
($a=0.320$ instead of $0.291 \pm 0.013$) and intercept ($b=2.249$
instead of $2.231 \pm 0.006$).  These are $2-3\, \sigma$ statistical
deviations. The steeper slope is visually evident in the bottom left
panel of Figure \ref{fig:tf_sigm}.  The model and observed relations
cross, but at a mass below $M_{*0}$ where the intercept is defined.
Relative to the constant--$m_d$ models, the total $\chi^2$ declines
from 40 to 35 in the AC model and from 13 to 10 in the no--AC model,
but since adding $\sigma_m$ as an adjustable parameter decreases the
number of degrees of freedom from three to two, the reduced $\chi^2$
actually goes up in each case.

% Note however, that despite the apparently different fit parameters,
% the model and observed distributions of points $M_*$ and $V_{2.2}$ are
% not too different (cf. bottom left panel of Fig.~\ref{fig:tf_sigm}).
% This emphasizes a strong degeneracy between the inferred fit parameters.

The efficiency of gas cooling in galactic halos or gas ejection or
heating by feedback could plausibly depend on galaxy mass and density.
Accordingly, the typical disk mass fraction may vary with these
parameters, and \citet{pizagno_etal05} argued for lower average $m_d$
values at low stellar masses.  Next, we consider a model where the
central value $\bar{m}_d$ scales with the stellar mass as
\begin{equation}
  \bar{m}_d = \bar{m}_{d,0} \, \left({M_* \over 10^{10.5}\, \Msun}\right)^{p_1}.
  \label{eq:ms}
\end{equation}
We have run a grid of models, varying the normalization
$\bar{m}_{d,0}$, the log-normal width $\sigma_m$, and the exponent
$p_1$, to find the minimum $\chi^2$.  Figure~\ref{fig:tf_ms} shows the
best-fit models for the cases with and without the AC effect, for
which the exponents $p_1 = 0.3$ and $p_1 = 0.1$, respectively.  By
boosting the circular velocities of low $M_*$ galaxies, the scaling
with stellar mass significantly improves the slope and normalization
of the TF relation for models with AC (both are now within $1\,
\sigma$ of the data).  However, the predicted anti-correlation of the
residuals is still inconsistent with the data, as is evident
from the lower right panel of Figure~\ref{fig:tf_ms}.  The no--AC
model with $M_*$ scaling is consistent with the observed slope,
intercept, and scatter, but it is still marginally inconsistent with
the residual correlation ($r = -0.17$ vs. $0.23 \pm 0.14$).

These models now have three adjustable parameters ($\bar{m}_{d,0}$,
$p_1$, $\sigma_m$, with $\sigma_c$ fixed by theory) and the degeneracy
among the best-fit values is substantial.  Within the limit
$\Delta\chi^2 = 1$, the range of allowed model parameters is $\Delta
p_1 \approx 0.1$, $\Delta \sigma_m \approx 0.2-0.3$.  Our inferred
parameters should therefore be taken as illustrating a preferred, but
by no means unique, trend.  Note that even three adjustable parameters
do not allow us to achieve an acceptable match to the four TF
parameters ($a$, $b$, $\sigma$, $r$) or to the four FP parameters.
Roughly speaking, $\bar{m}_{d,0}$ affects the intercept (high $m_d$ --
low $b$) and residual correlation (high $m_d$ -- strongly
anti-correlated residuals), $\sigma_m$ affects the scatter (high
$\sigma_m$ -- high scatter), and $p_1$ affects the slope (high $p_1$ --
low $a$), with $\sigma_c$ producing additional scatter, but no element
in these models can remove the residual anti-correlation.

Disk galaxy models of \citet{firmani_avila-reese00} show a trend of
increasing disk contribution to the rotation curve with increasing
stellar surface density, which in their models arises from the
correlation of the star formation efficiency with the surface density.
In observational studies, \citet{zavala_etal03} and
\citet{pizagno_etal05} find that the ratio of halo mass to stellar
mass within the luminous regions of disks correlates strongly with
disk surface density.  Motivated in part by these results, we consider
a set of models where the central value $\bar{m}_d$ scales with the
stellar surface density, $\Sigma_* \equiv M_* \, R_d^{-2}$:
\begin{equation}
  \bar{m}_d = \bar{m}_{d,0} \, 
    \left({M_* \, R_d^{-2} \over 10^{9.2}\, \Msun\, {\rm kpc}^{-2}}\right)^{p_2}.
  \label{eq:sden}
\end{equation}
The best-fit scalings for $p_2$ are 0.65 (AC) and 0.2 (no AC),
respectively.  Figure~\ref{fig:tf_sden} shows that such models
reproduce both the TF and FP relations fairly well, even including the
AC effect.  Given the mass-size relation (eq. [\ref{eq:r-m}]),
distributions (\ref{eq:ms}) and (\ref{eq:sden}) are internally
consistent: $\Sigma_* \propto M_*^{1/2}$ and $p_2 \approx 2 p_1$ (see
Table \ref{tab:tf}).  However, tying $\bar{m}_d$ to $\Sigma_*$ leads
to substantially better agreement with the data, greatly reducing the
anti-correlation of TF residuals.  The model without AC is consistent
with the observed slope, intercept, and scatter at the $1\, \sigma$
level, while its predicted residual anti-correlation remains $\sim 2\,
\sigma$ discrepant, $r = -0.07$ vs. $0.23 \pm 0.14$.  The FP
parameters are consistent at the $1\, \sigma$ level, except for the
radius coefficient, which is $a_2 = 0.022$ vs. $0.085 \pm 0.054$.  The
AC model remains notably discrepant with the TF residual correlation,
$r = -0.18$, and the FP radius scaling, $a_2 = -0.022$, but these
discrepancies are much smaller than in any of our previous models, and
other parameters agree within $1\, \sigma$.  If we minimize $\chi^2$
with respect to the FP parameters instead of the TF parameters, we
obtain somewhat stronger density scalings; these higher $p_2$ models
are also listed in Tables \ref{tab:tf} and \ref{tab:fp}.

Surface density scaling improves the TF and FP fits because it places
higher surface density galaxies in lower mass halos.  The
anti-correlation between disk-induced rotation velocity and
halo-induced rotation velocity counters the trend for more compact
disks to rotate faster -- in effect, there is no correlation between
radius residual and velocity residual because two correlations of
opposite sign cancel each other.  While this balance of competing
effects may seem coincidental, a correlation of $\bar{m}_d$ with
$\Sigma_*$ seems eminently reasonable on physical grounds.  Cooling
and star formation should be more efficient for denser baryon
distributions, and supernova feedback is probably less efficient at
ejecting gas from higher surface density disks.  In fact, the causal
arrow may point in the other direction -- high $\Sigma_*$ disks may
have higher surface density largely because they have higher $m_d$.
Finally, since we define $m_d$ in terms of disk stellar mass, a trend
of $m_d$ will arise automatically from the larger gas fractions of low
surface brightness galaxies.  This trend is driven physically by the
correlation of star formation rate with gas surface density
\citep{schmidt59, kennicutt98}.

Models without AC constitute a variation of our standard models (with
AC).  As another variation, we consider models with a light stellar
IMF.  We reduce stellar mass-to-light ratios (eq. [\ref{eq:mtol}]) by
0.15 dex relative to the Kroupa values.  The less massive stellar disks
now require more massive dark matter halos to reproduce the observed
$V_{2.2}$, and because adiabatic contraction has less impact, the
best-fit values of $m_d$ decrease by a factor substantially larger
than $10^{0.15}$.  Figure \ref{fig:tf_imf1} shows the best-fit model
with AC and $\bar{m}_d \propto \Sigma_*^{0.4}$.  The normalization in
equation (\ref{eq:sden}) drops from $\bar{m}_{d,0} = 0.1$ to
$\bar{m}_{d,0} = 0.04$.  This model has no TF residual
anti-correlation and the right amount of intrinsic scatter, although
the slope and intercept of the TF relation depart from the observed
values more than in our standard Kroupa--IMF model.  This model is
also statistically consistent with the observed FP relation,
reproducing its mass coefficient, intercept, and scatter, and lying
within $1.2\, \sigma$ of the radius coefficient $a_2$.  Although we
have no additional evidence to justify using the reduced mass-to-light
ratios, this alternative provides a good description of our galaxy
sample.  If the population synthesis models themselves are slightly
inaccurate, in the sense of predicting $M_*/L_i$ values that are too
high for a given $g-r$ color, then part of the desired 0.15--dex
reduction might arise even without a change to the IMF.

\begin{figure*}[t]
\vspace{-0.5cm}
\centerline{\epsfysize3.5truein \epsffile{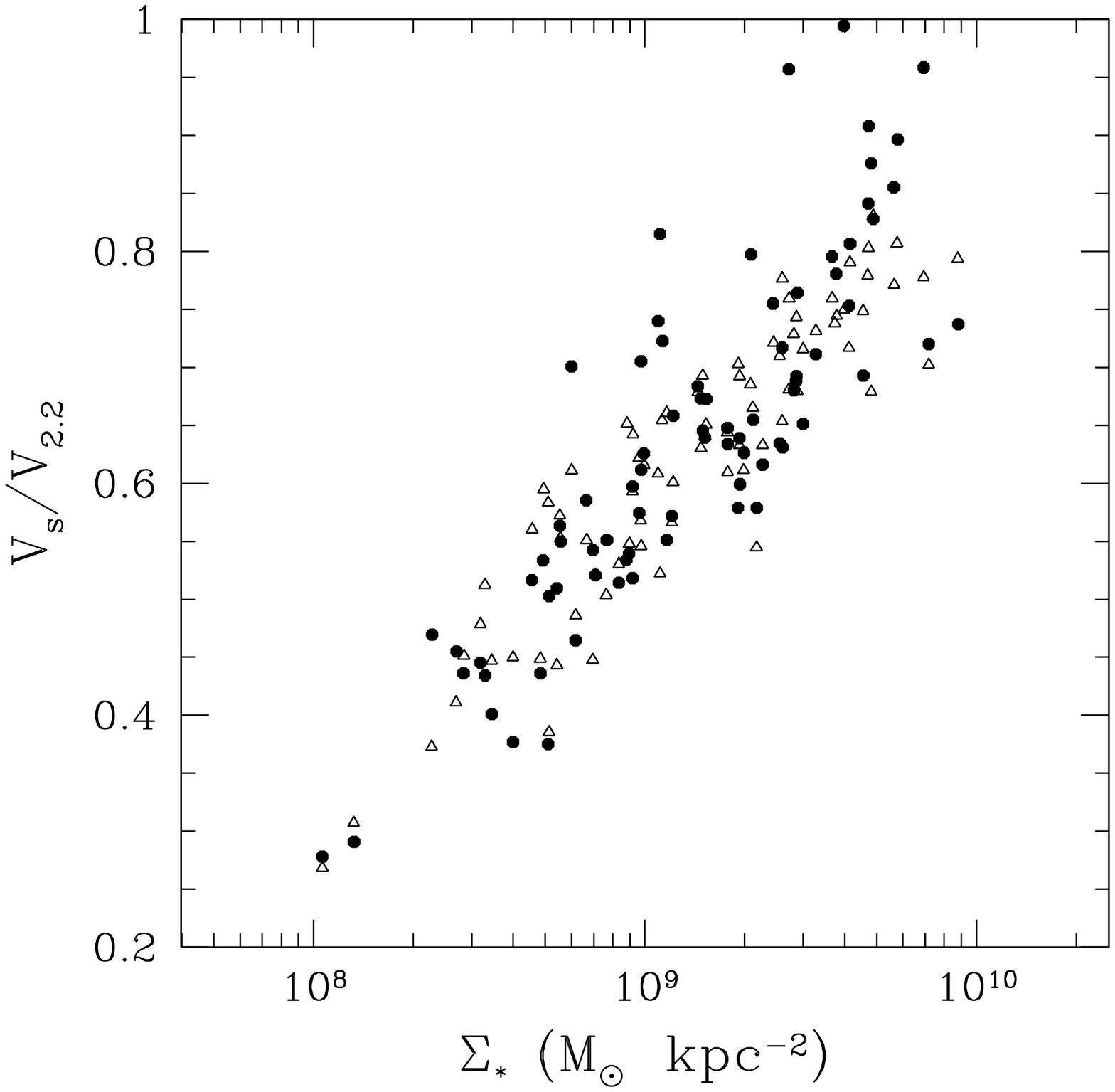}
            \epsfysize3.5truein \epsffile{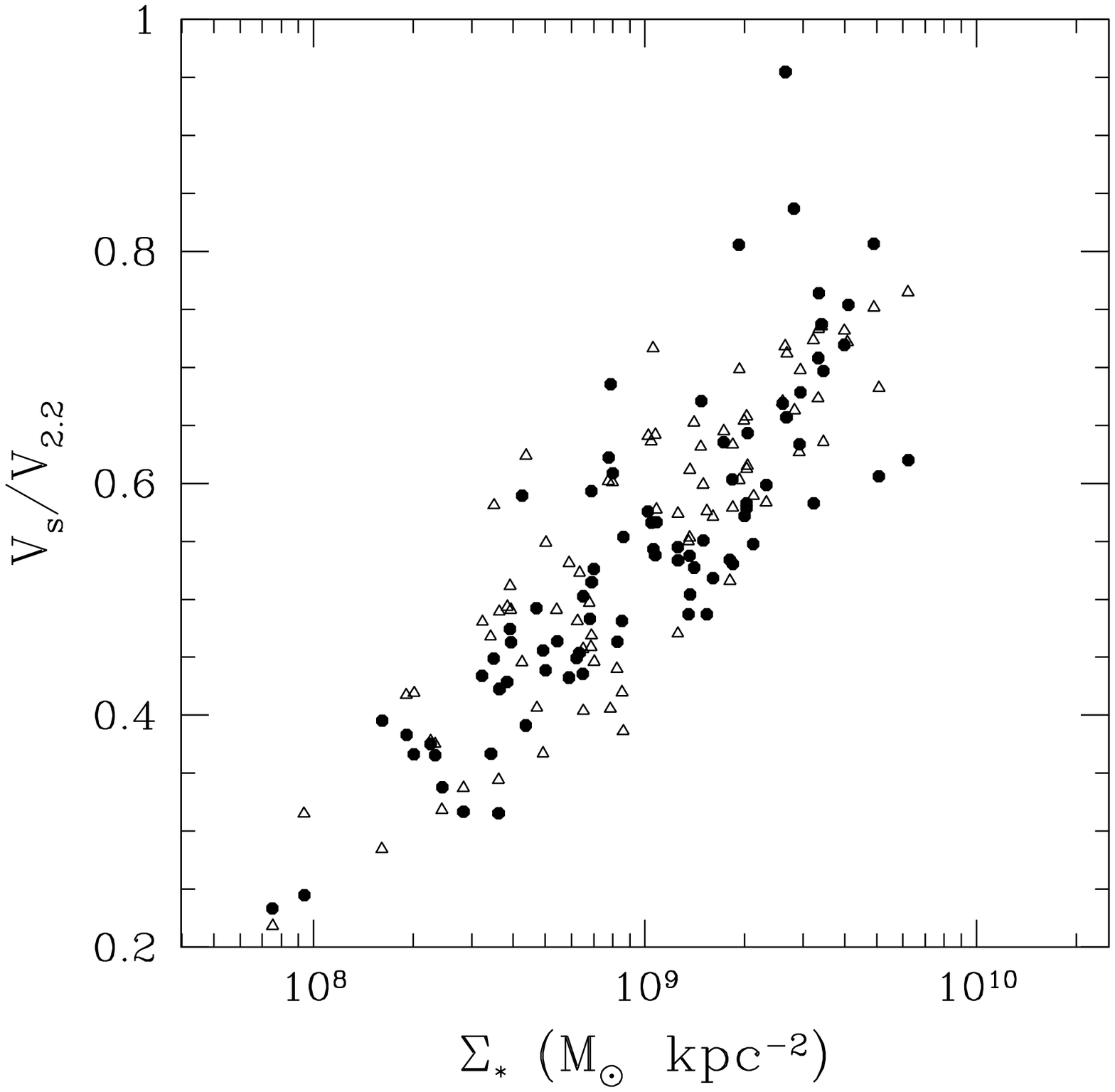}}
\vspace{-0.3cm}
\caption{Contribution of stars to the circular velocity at $2.2 R_d$
  vs. stellar surface density.  Filled circles are the data, using the
  SED-based estimates of the stellar mass, triangles are a realization
  of our best-fit model with $\bar{m}_d \propto \Sigma_*^{p_2}$,
  including the AC effect.  {\it Left panel:} Kroupa IMF, {\it right
  panel:} light IMF.}
  \label{fig:vs}
\end{figure*}

One of the persistent issues in disk galaxy studies is the role of
disk gravity in determining the rotation curve.  In ``maximal disk''
models, the disk typically contributes $75\% - 95\%$ of the rotational
support at $2.2 R_d$ \citep{sackett97}.  Solid points in the left
panel of Figure~\ref{fig:vs} show the ratio of the circular velocity
from the stellar disk to the measured circular velocity $V_{2.2}$ for
our observed sample, assuming a Kroupa IMF (similar to Fig. 7 of
\citealt{pizagno_etal05}).  The contribution of stars to the circular
velocity at $2.2 R_d$ increases monotonically with the stellar surface
density.  Most of the galaxies with $\Sigma_* > 3\times 10^9\, \Msun\,
{\rm kpc}^{-2}$ are maximal disk by Sackett's criterion, while the
less dense disks are dark matter-dominated.\footnotemark
\footnotetext{When using the number $3\times 10^9\, \Msun\, {\rm
kpc}^{-2}$, one should keep in mind our definition of the surface
density, $\Sigma_* \equiv M_* \, R_d^{-2}$, not containing any factors
of $2\pi$.}
Open points show the same ratio for the model realization shown in the
lower panel of Figure~\ref{fig:tf_sden}, with $\bar{m}_d \propto
\Sigma_*^{0.65}$ and AC.  The model shows the same trend as the data,
except that there are essentially no model galaxies with $V_s/V_{2.2}
> 0.8$.  Ratios higher than this imply such small halos (relative to
the galaxy baryon mass) that our model never predicts them.  The right
panel of Figure~\ref{fig:vs} shows the same results for the light IMF.
Note that the observational data points also change in this case,
because the value of $M_*$ inferred for each galaxy is lower and the
values of $\Sigma_*$ and $V_s$ are therefore lower as well.  Only two
observed galaxies have $V_s/V_{2.2} > 0.8$ in this case, while the
rest have sub-maximal disks.  The correlation of the ratio
$V_s/V_{2.2}$ with $\Sigma_*$ is stronger than with $M_*$ or other
available galaxy parameters.  This reinforces the assumption of our
models that the surface density is a major factor regulating the
stellar mass fraction of disk galaxies.

\begin{figure}[t]
\vspace{-0.5cm}
\centerline{\epsfysize3.5truein \epsffile{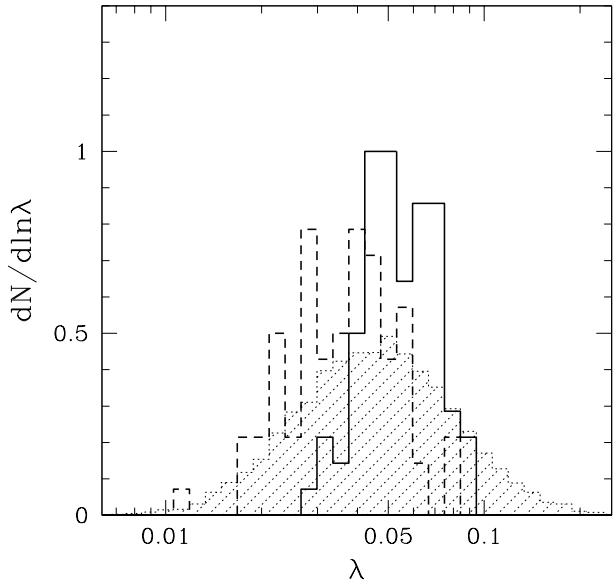}}
\vspace{-0.5cm}
\caption{Distribution of disk angular momentum parameter required by
  our best-fit models with $\bar{m}_d \propto \Sigma_*^{p_2}$ with AC
  ({\it solid histogram}) and without AC ({\it dashed histogram}).
  Dotted hatched histogram shows the $P(\lambda)$ distribution of dark
  matter halos as in Fig. \protect\ref{fig:lam}.}
  \label{fig:lam2}
\end{figure}

Figure~\ref{fig:lam2} shows the distributions of the disk angular
momentum parameter for the two models of Figure~\ref{fig:tf_sden}.
These $\lambda_d$ histograms are qualitatively similar to those for
the constant--$m_d$ models shown in Figure~\ref{fig:lam}.  The typical
$\lambda_d$ values again increase with the typical values of $m_d$.
The mode values are $\bar{\lambda}_d \approx 0.054$ for the AC model
and $\bar{\lambda}_d \approx 0.036$ for the no--AC model, compared to
$\bar{\lambda} \approx 0.045$ for dark matter halos.  For visual
clarity, we have omitted the light--IMF model from
Figure~\ref{fig:lam2}, but its $\lambda_d$ histogram is intermediate
between the other two, with $\bar{\lambda}_d \approx 0.040$.  All
three model distributions are narrower than the corresponding
$P(\lambda)$ for dark matter halos: $\sigma_{\lambda d} \approx 0.25$
(AC), 0.39 (no AC), 0.34 (light IMF) vs. $\sigma_\lambda \approx
0.56$.  Furthermore, our procedure should overestimate the true width
of the $\lambda_d$ histogram because we choose a random value of $m_d$
from $P(m_d | M_*, R_d)$, thus adding artificial scatter to the
$\lambda_d$ values given the $R_d$ constraints.  (We do not expect to
draw the correct value of $m_d$ on a galaxy-by-galaxy basis, even if
our model incorporates the correct spread of $m_d$ values.)  To test
the magnitude of this effect, we have applied our procedure to a
realization of the AC model, choosing new, random values of $m_d$ and
$c$ for each model galaxy.  The input $\lambda_d$ distribution has a
standard deviation $\sigma_{\lambda d} = 0.25$, while the distribution
inferred by fitting the model data points has $\sigma_{\lambda d} =
0.34$, nearly 40\% larger.  The mean value $\bar{\lambda}_d$ is
reproduced correctly.  Thus the $\lambda_d$ distributions shown in
Figure~\ref{fig:lam2} place strong constraints on the allowed values
of the angular momentum parameter of disk-dominated galaxies.  These
distributions are quite different from the $\lambda$ distribution of
$N$-body halos, implying that only a subset of halos form disk
galaxies like those in our sample, or that disk baryons have
systematically different angular momenta from the dark matter.

% \citet{dutton_etal05,dutton_etal06} have reported a similar study of
% constraining the models of disk galaxy formation using a larger but more 
% heterogeneous sample of local galaxies.  In agreement with our data,
% their sample shows a very weak (but negative) correlation of the TF
% residuals, consistent with no correlation.  Unlike our approach,
% Dutton et al. do not use the observed disk sizes and luminosities in
% construction of the models, but instead fit the size-luminosity
% relation simultaneously with the luminosity-velocity relation.  They
% find that the best-fit model requires the stellar fraction to increase
% with galaxy mass as $\bar{m}_d \propto M^{0.18}$ (or correspondingly,
% with stellar mass as $M_*^{0.15}$), as do our models (see Table
% \ref{tab:tf}).  Dutton et al. attribute this mass dependence to the
% density threshold for star formation.  Contrary to our results, Dutton
% et al. find that no model including the AC effect can fit
% simultaneously the size-luminosity and the luminosity-velocity
% relations.  Since we find that only the model with $\bar{m}_d \propto
% \Sigma_*^{p_2}$ produces adequate fit to our data, and Dutton et
% al. did not consider such a model, the density-dependence of the disk
% fraction may be key to bringing the halo contraction effect in
% agreement with the data.  Note however, that the best-fit models with
% the AC effect still over-predict the number of disk galaxies, as we
% discuss in \S\ref{sec:lf}.

\citet[][hereafter DVDC]{dutton_etal06} have recently carried out an
investigation that is similar to ours in many respects, though in
spirit theirs is more a theoretical modeling study constrained by data
while ours is an empirical inference study informed by theory.  
% DVDC use Courteau et al.'s (2006) compilation of the Mathewson et
% al. (1992), Dale et al. (1999), and Courteau et al. (2000) $I$-band TF
% data sets.
The observational results from their large compilation of three
$I$-band TF data sets ($\sim$ 1300 galaxies with H$\alpha$ rotation
curves) agree well with those from our smaller, homogeneous data set,
which is reassuring given the numerous differences in sample
selection, observations, analysis methods, and typical galaxy
distances (which strongly affect the observational error budget
because of peculiar velocities).  In particular (see Fig. 3 of DVDC)
we find similar TF slope and intercept and similar lack of correlation
between TF residual and radius residual; our galaxies have slightly
larger average scale lengths at fixed luminosity, which is plausibly
an effect of our stricter limit on bulge-to-disk ratio. 
%
% DVDC's estimated intrinsic scatter of $\log V_{2.2}$ at fixed $I$-band
% luminosity, inferred by subtracting typical observational errors in
% quadrature from the total scatter, is similar to our
% maximum-likelihood estimate of the intrinsic scatter at fixed {\it
% stellar mass}.  Although our approaches differ in that DVDC include
% the galaxy-to-galaxy scatter in $I$-band mass-to-light ratios and do
% not include distance uncertainties in their observational error budget
% (while we include the latter but not the former), the two effects must
% approximately balance each other.
%
DVDC's estimated intrinsic scatter of 0.052 dex in $\log V_{2.2}$ at
fixed $I$-band luminosity, inferred by subtracting observational
errors in quadrature from the total scatter, is only marginally larger
than our maximum-likelihood estimate of the intrinsic scatter at fixed
{\it stellar mass}, $\sigma = 0.049 \pm 0.007$ dex.  This similarity
is somewhat surprising, since the scatter at fixed luminosity should
be increased by galaxy-to-galaxy scatter in $I$-band mass-to-light
ratios.  Furthermore, if DVDC included distance uncertainties in their
observational error budget, their estimate of the intrinsic scatter
would decrease, coming closer to our value or perhaps even falling
below it.

There are many differences of detail in our theoretical modeling
methods: we model relations involving estimated stellar masses while
DVDC use $I$-band luminosities; we infer $\lambda_d$ distributions by
fitting individual $R_d$ values while DVDC assume a log-normal
$\lambda_d$ distribution of adjustable width; and we adopt a
\cite{kroupa01} IMF as our standard while DVDC take Bell \& de Jong's
(2001) ``diet Salpeter'' IMF as fiducial.  Despite these differences,
our qualitative theoretical conclusions are largely concordant to the
extent that they can be compared.\footnote{As we were already
finalizing our manuscript when the DVDC preprint became available, we
have not carried out a comprehensive comparison of results.}  Both
studies infer a narrow distribution of $\lambda_d$ relative to N-body
predictions for the halo $\lambda$ distribution.  DVDC do not consider
models with $m_d$ connected to $\Sigma_*$.  However, their surface
density threshold for star formation has an effect of the same sign
(since they define $m_d$ in terms of baryonic mass while we define it
in terms of stellar mass), and they find that this threshold plays an
important role in suppressing TF residual correlations. We find that
scatter in $m_d$ is required to reproduce the observed TF scatter,
while DVDC do not.  Given the similar observational estimates of the
intrinsic scatter, this difference probably arises largely because
DVDC incorporate scatter in $I$-band stellar mass-to-light ratios
while we use $g-r$ color to estimate $M/L_i$ and assume no scatter
about the mean relation.  DVDC emphasize the galaxy luminosity
function as an additional constraint on models, and we will take up
this point ourselves in \S\ref{sec:lf} below (see Figure
\ref{fig:mf}).  We concur with DVDC that omitting adiabatic
contraction, adopting a lighter IMF, or lowering halo concentrations
(see \S\ref{sec:lf}) all help to improve agreement with the observed
TF relation (including residual correlations) and to reduce the
tension between the TF relation and the galaxy stellar mass function
(or luminosity function).  We are more skeptical about the physical
plausibility of omitting (or, in DVDC's favored model, reversing) the
effects of adiabatic contraction, for the reasons cited in
\S\ref{sec:intro}, so we lean towards some combination of light IMF
and lower halo concentration as a more likely explanation.

\subsection{Individual galaxy (``exact'') fit models}

A common approach in modeling the TF relation has been to choose a
distribution $P(\lambda_d)$ {\it a priori} \citep{kauffmann_etal93,
dalcanton_etal97, mo_etal98, vandenbosch00}.  In the previous
sections, we have modified this approach by choosing $\lambda_d$ to
match the observed disk scale length $R_d$ on a galaxy-by-galaxy basis.
In doing so, we have derived an empirical estimate of $P(\lambda_d)$,
given the model distribution $P(m_d | M_*,R_d)$.  We can go one step
further and choose $m_d$ for each galaxy to exactly reproduce its
observed velocity $V_{2.2}$.  We still assume that the dark matter
halos are described by the NFW model with a concentration parameter
dependent on $M$.  Models constructed in this way reproduce the
observed $M_*$, $R_d$, and $V_{2.2}$ distributions by construction,
and they yield $P(m_d)$ as a result.

\citet{jimenez_etal03} follow this approach to estimate the disk mass
fractions and angular momentum parameters from a large sample of
H$\alpha$ rotation curves, assembled from various sources.  They
consider cuspy NFW models and cored pseudo-isothermal spheres for the
dark halos and apply the standard model of adiabatic contraction
\citep[e.g.,][]{blumenthal_etal86}.  In contrast to our procedure,
they do not use stellar mass-to-light ratios estimated from galaxy
colors but instead treat $M_*/L$ as an additional fitting parameter
for each galaxy.

There are two reasons why we do not adopt this ``exact fit'' method as
our main approach.  First, we choose the concentration parameter at
random from the distribution $P(c)$, and since the inferred disk mass
fraction is sensitive to the value of $c$, the $m_d$ values are not
accurate on a galaxy-by-galaxy basis.  Second, observational errors on
$M_*$, $R_d$, and $V_{2.2}$ introduce substantial uncertainty in the
derived values of $m_d$, even if the assumed concentration is correct.
(The same effect influences the $\lambda_d$ values in our standard
approach, but much more weakly.)  Despite these shortcomings, it is
interesting to compare the $P(m_d)$ estimated by this ``exact fit''
approach to the distribution assumed in the best-fit models of section
\ref{sec:mdscatter}.

\begin{figure*}[t]
\vspace{-0.5cm}
\centerline{\epsfysize3.5truein \epsffile{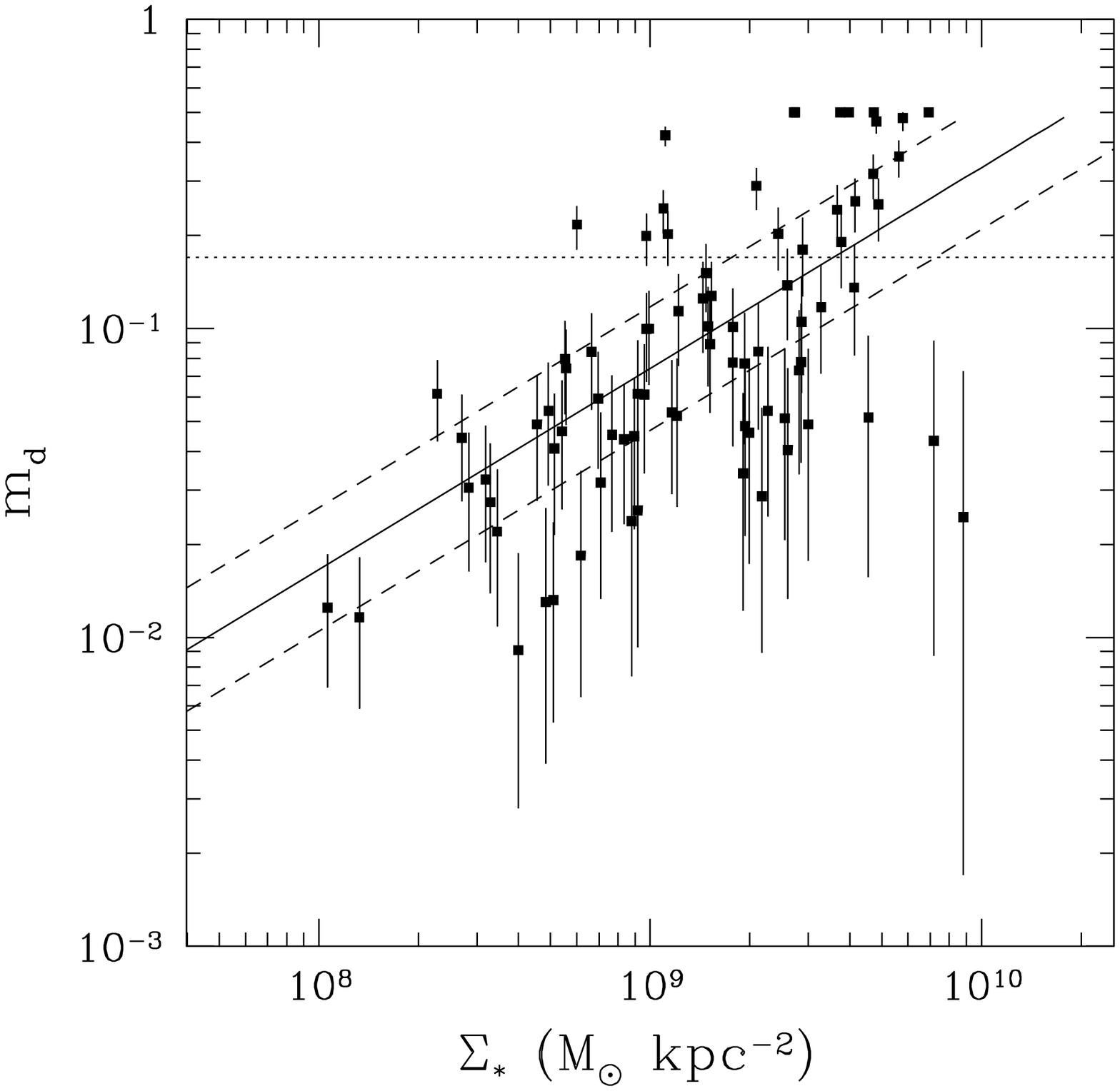}
            \epsfysize3.5truein \epsffile{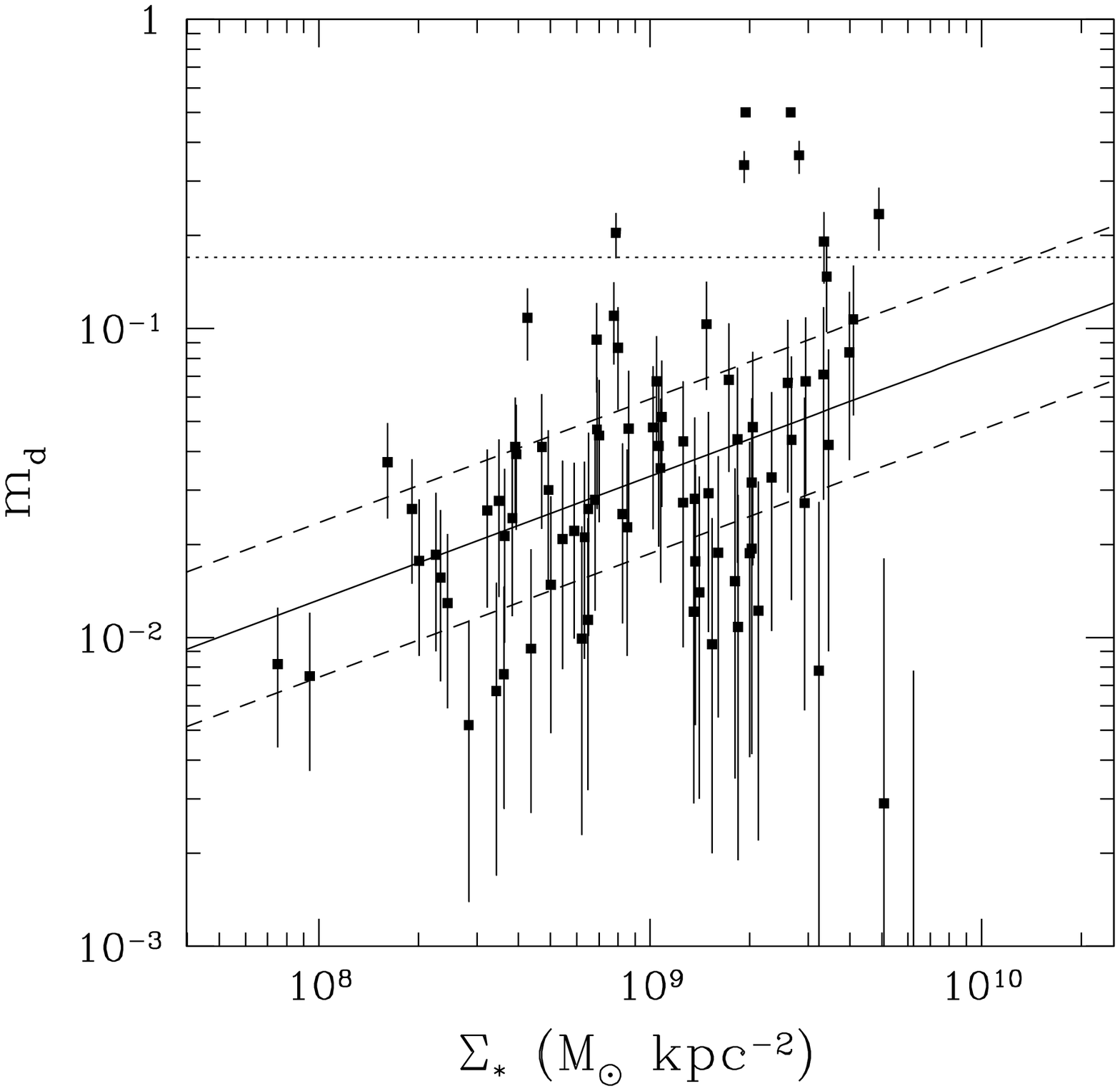}}
\vspace{-0.3cm}
\caption{Disk mass fractions that reproduce the observed rotation velocities
  exactly for models with AC and two stellar IMFs: Kroupa IMF ({\it
  left panel}), light IMF ({\it right panel}).  Vertical error bars
  illustrate the uncertainty associated with the scatter of halo
  concentrations.  The end points correspond to $\log{c} \pm
  \sigma_c$, with $\sigma_c=0.14$.  The value of $m_d$ is not allowed
  to exceed 0.5.  Lines show the mean trend ({\it solid}) and $1\,
  \sigma$ scatter ({\it dashed}) of the best-fit models with AC
  (eq. [\protect\ref{eq:sden}]) for each IMF.}
  \label{fig:mdms}
\end{figure*}

% \begin{figure}[t]
% \vspace{-0.5cm}
% \centerline{\epsfysize4truein \epsffile{f13.eps}}
% \vspace{-0.5cm}
% \caption{Distribution of disk angular momentum parameter required to
%   reproduce the data exactly.  The models include the effect of
%   adiabatic contraction of dark halos.}
%   \label{fig:lmatch}
% \end{figure}

Figure~\ref{fig:mdms} plots the inferred values of $m_d$ against
$\Sigma_*$, assuming AC and a Kroupa IMF (left panel) or a light IMF
(right panel).  Points show the $m_d$ value determined for a halo
concentration $c = \bar{c}(M)$.  Error bars show the effect of raising
or lowering the log of the concentration by $\pm \sigma_c$ (0.14 dex).
For higher $c$, the total halo mass must be lower to produce the same
$V_{2.2}$, and the value of $m_d = M_*/M$ is consequently higher.  We
have also tried varying the values of $M_*$, $R_d$, and $V_{2.2}$ by
their $1\, \sigma$ observational errors, and these variations on their
own produce an $m_d$ spread similar to that shown by the error bars in
Figure~\ref{fig:mdms}.

Lines in the left panel of Figure~\ref{fig:mdms} show the mean trend
and $1\, \sigma$ scatter of our best-fit Kroupa--IMF model from
\S\ref{sec:mdscatter}.  The trend of the ``exact fit'' points is
qualitatively similar to that of the model.  However, there are a
number of galaxies that cannot be fit with an $m_d$ value below the
universal baryon fraction of 0.17, for any value of $c$ in the
expected range.  (We have arbitrarily imposed an upper limit of 0.5 on
$m_d$ in Figure~\ref{fig:mdms}.)  These are the ``maximal disk''
galaxies from Figure~\ref{fig:vs}, which cannot be fit by an
adiabatically contracted NFW halo with typical concentration.  If we
omit adiabatic contraction, some but not all of these points move to
$m_d < 0.17$.  For the light IMF, only four galaxies require $m_d >
0.17$ for a concentration in the expected range.  The smaller number of
physically peculiar results in this case is a circumstantial argument
in favor of the light IMF.

\begin{figure}[t]
\vspace{-0.5cm}
\centerline{\epsfysize3.5truein \epsffile{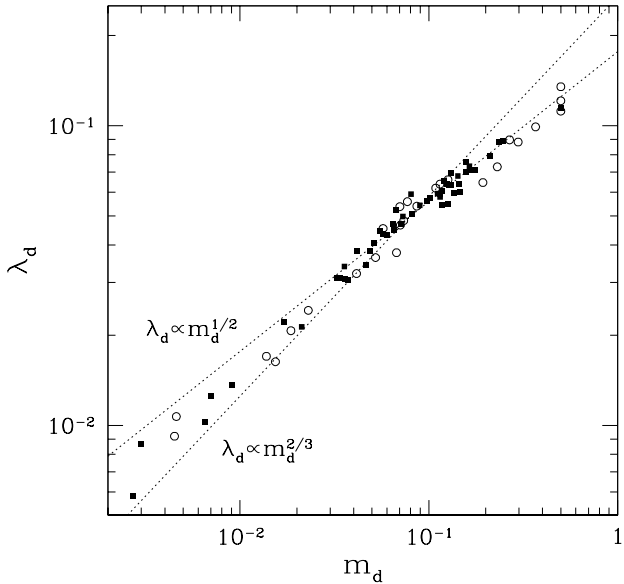}}
\vspace{-0.3cm}
\caption{Scatter in the inferred ``exact-fit'' values of $m_d$ and
  $\lambda_d$ resulting from observational errors in $M_*$, $R_d$,
  $V_{2.2}$ and from scatter of halo concentrations.  The dotted lines
  show two power-law trends, $\lambda \propto m_d^{1/2}$ and $\lambda
  \propto m_d^{2/3}$, respectively.  True values were set to $m_d =
  0.08$ and $\lambda_d = 0.05$ (at the intersection of the dotted
  lines).  Solid squares and open circles correspond to galaxies with
  $\Sigma_*$ above and below $10^{9.2}\, \Msun\, {\rm kpc}^{-2}$,
  respectively.}
  \label{fig:ldmd_test}
\end{figure}

It is tempting to go further and examine, for example, the correlation
between $m_d$ and $\lambda_d$ \citep[see][]{jimenez_etal03}.
Unfortunately, we consider correlations derived in this way to be
unreliable.  Adopting an incorrect value of $c$ for an individual
galaxy leads to an incorrect $m_d$, and the inferred value of
$\lambda_d$ for that galaxy adjusts in a correlated way to reproduce
the observed disk scale length $R_d$.  If we take an input model that
has no correlation between $m_d$ and $\lambda_d$, we find that this
effect produces a strong, spurious correlation between the ``exact
fit'' values of $m_d$ and $\lambda_d$.  A similar spurious correlation
arises from observational errors, as we illustrate in Figure
\ref{fig:ldmd_test}.  For this test, we create an artificial input
sample by choosing the constant values of $\lambda_d = 0.05$ and $m_d
= 0.08$ for all observed galaxies and recalculating $R_d$ and
$V_{2.2}$ accordingly, while keeping the observed $M_*$ and the mean
halo concentration $\bar{c}(M)$.  Then we generate a realization of
the new sample with added random ``measurement'' errors of $M_*$,
$R_d$, $V_{2.2}$ and perform the ``exact fit'', assuming a halo
concentration drawn from $P(c)$.  The inferred values of $m_d$ and
$\lambda_d$ scatter along the lines of $\lambda_d \propto m_d^{1/2}$
at high $m_d$ and $\lambda_d \propto m_d^{2/3}$ at low $m_d$,
displaying a spurious correlation.  It might be possible to
investigate the correlation of $m_d$ and $\lambda_d$ by fitting
resolved rotation curves and thus deriving the halo concentration for each
galaxy, but even in this case one should be cautious of the impact of
observational errors.

We have also investigated the dependence of inferred values of $m_d$
and $\lambda_d$ on $\Sigma_*$.  The underlying question is: Do low
surface brightness galaxies have low density because of low $m_d$
(feedback or inefficient star formation) or because of high
$\lambda_d$ (high angular momentum)?  We use the radius residual,
$\Delta R_d$, relative to the mean $R_d - M_*$ relation as our measure
of surface density variation at fixed $M_*$.  For our best-fit AC,
Kroupa--IMF model we find only a weak anti-correlation between $\Delta
R_d$ and $m_d$, but we do find a significant positive correlation
between $\Delta R_d$ and $\lambda_d$ (correlation coefficient is
0.63).  Thus in our best-fit model, the low-density galaxies tend to
have both lower $m_d$ (a feature built into the model by the
$m_d-\Sigma_*$ correlation) and higher $\lambda_d$, but the
$\lambda_d$ variations dominate the spread of surface density at fixed
$M_*$.  We are unable to confirm or reject this trend using our
``exact fit'' procedure, because any correlation is washed out by
observational uncertainties and by scatter of halo concentrations.
Figure~\ref{fig:ldmd_test} shows that the inferred values of $m_d$ and
$\lambda_d$ are scattered similarly for the high and low surface
density galaxies.  Even when we create another artificial input sample
with an encoded anti-correlation of $\lambda_d$ and $\Sigma_*$, the
recovered values of $\lambda_d$ are completely uncorrelated with
$\Sigma_*$.

% The distribution is broadly consistent with the best-fit scaling
% relations (eq. [\ref{eq:sden}]), but with a much larger dispersion at
% a given stellar mass.
% Note also the large uncertainties in the inferred $m_d$ due to
% the scatter of halo concentrations.  Moreover, adding measurement
% errors (in $V_{2.2}$, $M_*$, and less so in $R_d$) produces a similar
% amount of scatter in the resulting $m_d$.  Thus, this large dispersion
% may appear by construction, as a result of the observational errors
% convolved with an intrinsically narrow distribution $P(m_d)$.

% The inferred disk angular momentum parameter $\lambda_d$ is higher than the
% mean dark matter $\lambda$.  This is consistent with previous studies,
% which find that disk galaxies form from high-angular momentum gas.
%
% Figure~\ref{fig:lmatch} shows the angular momentum parameters derived
% for the ``exact'' galaxy models.
%
% This distribution (not shown for brevity) is shifted towards higher
% $\lambda_d$ than that in Figure~\ref{fig:lam2} ($\bar{\lambda}_d =
% 0.059$) and is significantly broader ($\sigma_{\lambda d} = 0.41$).
% \citet{jimenez_etal03} found the distribution $P(\lambda_d)$ biased
% towards lower values than those of dark matter, in apparent
% disagreement with our best-fit models with the AC effect.  However,
% note that their sample includes some Sa type galaxies with a
% significant bulge contribution, which could lower the effective values
% of $m_d$ and $\lambda_d$ compared to our disk-dominated sample.

\section{Additional constraints}
  \label{sec:lf}

So far, we have constrained the distribution of disk mass fractions
using only the TF relation and its residuals, or the FP relation.  In
this section, we compare the best-fit models from
\S\ref{sec:mdscatter} to additional observational constraints from the
stellar mass function of galaxies and from the extended mass profiles
around galaxies probed by weak lensing and satellite dynamics.  For
brevity, we will refer to the three models shown in Figures
\ref{fig:tf_sden} and \ref{fig:tf_imf1}, all of which incorporate
$\bar{m}_d \propto \Sigma_*^{p_2}$, as the AC, no--AC, and light--IMF
models.

Given our best-fit $P(m_d | M_*, R_d)$, we can go from a cosmological
model prediction of the abundance of dark matter halos to a prediction
of the number density of disk galaxies as a function of stellar mass.
To make such predictions we use the halo mass function calculated by
the ellipsoidal excursion set method of \citet{sheth_tormen02}, which
accurately describes the results of cosmological $N$-body simulations
\citep[e.g.,][]{springel_etal05}.\footnotemark
\footnotetext{A study by \citet{jenkins_etal01} finds that the
excursion set method gives the best fit to the $N$-body mass function
when the halo mass is defined using a constant overdensity 180 times
the mean matter density.  We have adopted a higher virial overdensity,
$101/\Omega_m \approx 421$, which results in lower virial masses.  In
order to bring our halo masses to the same scale as the
\citet{sheth_tormen02} mass function, we assume a halo outer density
profile $\rho(r) \propto r^{-3}$ and find the radius where the mean
overdensity drops to 180.  This radius is 33\% larger than our adopted
$R_{\rm vir}$, and the enclosed mass is 15\% higher than our $M_{\rm
vir}$, for a concentration of galaxy-sized halos $c=12$.  Therefore,
we have shifted the \citet{sheth_tormen02} mass function by a factor
1.15 in halo mass.}
We adopt cosmological parameters of a $\Lambda$CDM model, $\Omega_{\rm
m} = 1 - \Omega_\Lambda = 0.24$, $\Omega_{\rm b}=0.041$,
$\sigma_8=0.74$, $n=1$, $H_0 = 70$ km s$^{-1}$ Mpc$^{-1}$, favored by
the WMAP 3-year data set and a variety of other cosmological
observables \citep{spergel_etal06}.  Note, however, that our adopted
$\bar{c}(M)$ is calibrated on simulations with higher $\sigma_8$, a
point we will return to shortly.  We use the models defined by
equation (\ref{eq:sden}) to convert the halo masses $M$ to stellar
masses $M_*$.  First, we use the mean mass-radius relation
(eq. [\ref{eq:r-m}]) to express the average surface density as a
function of stellar mass, $\Sigma_* \propto M_*^{0.52}$, and convert
the density-dependent model $\bar{m}_d(\Sigma_*)$ to the
mass-dependent $\bar{m}_d(M_*)$.  This procedure yields similar but
not identical functions to those given by equation (\ref{eq:ms}).
Then we integrate equation (\ref{eq:md}) to obtain the mean disk mass
fraction for galaxies of mass $M_*$: $\left< m_d \right>(M_*) \equiv
\int m_d P(m_d | M_*) d\log{m_d}$, with an upper limit $m_d = 0.15$.
Finally, we convert stellar mass to the mean halo mass, $\left< M
\right> = M_*/\left< m_d \right>(M_*)$, and shift the halo mass
function to a galaxy mass function, $dn_{\rm gal}(M_*)/d\ln{M_*}
\equiv dn_{\rm halo}({\left< M \right>})/d\ln{M}$.

% In the conversion from the halo mass to stellar mass, we have
% constrained our distribution of $m_d$ (eq. [\ref{eq:md}]) not to
% exceed the upper limit $m_{\rm max} = 0.15$.  This cutoff affects 
% the mean value of the disk fraction integrated over its distribution 
% function.  Setting $x \equiv \log{m_d}$, we find
% \begin{equation}
%   \langle\log{m_d}\rangle = \int_{-\infty}^{\log{m_{\rm max}}} x P(x) dx
%      = \log{\bar{m}_d} - {\sigma_m \over \sqrt{2\pi}} \,
%        {\exp{(-a^2)} \over 1 - {\rm erfc}(a)/2},
% \end{equation}
% where $a \equiv (\log{m_{\rm max}} - \log{\bar{m}_d})/\sqrt{2}
% \sigma_m$ and ${\rm erfc}(a)$ is the complementary error function.

Our calculation implicitly assumes that each dark matter halo hosts
one and only one disk galaxy, so it does not include the contribution of
satellite galaxies to the galaxy mass function.  However, theoretical
models and observational estimates from galaxy clustering and
galaxy-galaxy lensing all indicate that the fraction of galaxies that
are satellites in the mass range of our sample is low, between 10\%
and 20\% \citep{mandelbaum_etal05, zehavi_etal05, zheng_etal05}.

\begin{figure}[t]
\vspace{-0.5cm}
\centerline{\epsfysize3.5truein \epsffile{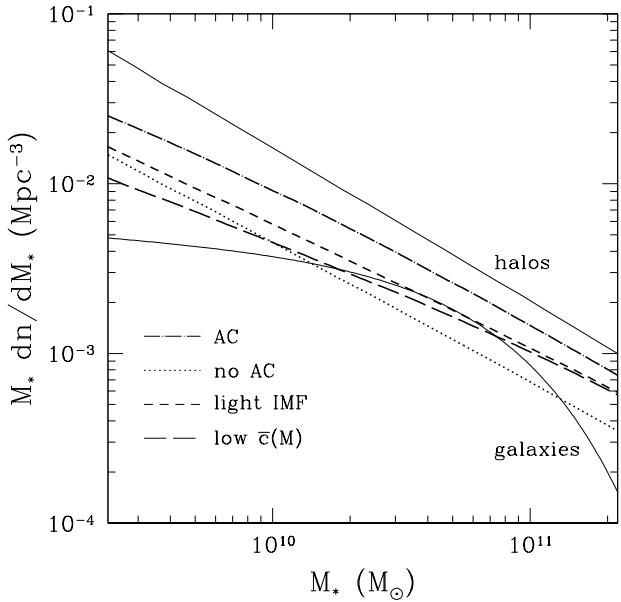}}
\vspace{-0.3cm}
\caption{Predicted mass functions of disk galaxies for the best-fit
  models with $\bar{m}_d \propto \Sigma_*^{p_2}$ with AC ({\it
  dot-dashed line}), without AC ({\it dotted line}), with AC and light
  stellar IMF ({\it dashed line}), and with low $\bar{c}(M)$ ({\it
  long-dashed line}).  These galaxy number densities are calculated
  from the abundance of dark matter halos in the concordance
  $\Lambda$CDM cosmology as a function of galaxy stellar mass, $M_* =
  m_d M$, with the model values of $m_d(M)$ in each case.  For
  comparison, top solid line (labeled ``halos'') shows the case if the
  halos converted all of their baryons into stars.  Bottom solid curve
  (labeled ``galaxies'') shows the observed mass function of galaxies
  in the local universe from \protect\citet{bell_etal03}.  The plotted
  range of masses corresponds to the range of rotation velocities
  $80-300$ km s$^{-1}$, covered by our observed sample.}
  \label{fig:mf}
\end{figure}

Figure~\ref{fig:mf} shows the theoretical halo mass function, the
galaxy stellar mass functions predicted by the AC, no--AC, and
light--IMF models, and the observational estimate of this mass
function by \citet{bell_etal03}, derived from the SDSS and 2MASS
surveys.  We have converted this estimate from Bell et al.'s
``diet-Salpeter'' IMF to a Kroupa IMF by multiplying their galaxy mass
parameters by $10^{-0.15}$.  The AC model substantially over-predicts
the observed mass function at all masses, even without accounting for
satellite galaxies.  The high values of $m_d$ required to match the TF
relation with a Kroupa IMF and AC puts galaxies of a given $M_*$ into
relatively low-mass, abundant halos, leading to a high number density.
This tension between the galaxy luminosity function and the
Tully-Fisher relation is a long-standing problem for CDM-based models
of galaxy formation \citep[e.g.,][]{kauffmann_etal93, cole_etal94,
somerville_primack99}.  Here we are essentially confirming this
problem, but removing some sources of uncertainty by using stellar
masses in place of luminosities and by applying TF constraints from a
well defined galaxy sample.  Of course, some of the halos in this
range ($V_{\rm vir} = 80-300$ km s$^{-1}$) might host galaxies with
significant bulges, which would not be in our TF sample -- about 50\%
of the galaxies in Pizagno et al.'s (2006, in preparation) full sample
satisfy the bulge-to-total $< 0.1$ criterion used here.  However, if
we lower the predicted mass function to account for the fraction of
galaxies with substantial bulges, then we should also lower the
observed mass function by the same factor, leaving the same gap
between them.  Reconciling the AC model with the \citet{bell_etal03}
estimate for a Kroupa IMF requires, instead, that the halos that do
not host disk-dominated galaxies have drastically lower values of
$m_d$, making them ``dim'' or ``dark''.

The dotted curve in Figure~\ref{fig:mf} shows the prediction of the
no--AC model, which lies below the \citet{bell_etal03} estimate near
the knee of the mass function but rises above it at low and high
masses.  We do not view the discrepancy above $10^{11}\, \Msun$ as a
serious problem, because a large fraction of galaxies in this mass
range are early-type systems, which could plausibly have $m_d$ values
lower by a factor $\sim 2$, and because massive halos have a larger
fraction of their baryon mass in satellite galaxies and intergalactic
gas, leaving less for the central galaxy.  The tension below $M_* \sim
2\times 10^{10}\, \Msun$ is more significant, since our sample extends
to $2\times 10^9\, \Msun$ and most galaxies in this mass range are
late types, but it is arguably within the systematic uncertainties of
the model and the data.  We therefore consider the no--AC model to be
an empirically acceptable way to reconcile the TF and mass function
constraints, albeit one that requires abandoning a seemingly well
established element of galaxy formation physics.

The dashed curve in Figure~\ref{fig:mf} shows the model with adiabatic
contraction and a light stellar IMF.  Rather than plot a second
version of the ``observed'' mass function, we have added 0.15 dex back
to the model values of $M_*$, thus allowing a direct comparison to the
original observational curve.  If changing the IMF simply lowered
$m_d$ values by 0.15 dex, then the predicted and observed mass
functions would shift by the same factor, and the agreement between
them would not improve.  However, because the light IMF reduces the
effect of disk gravity (both the direct contribution and the magnitude
of AC), it lowers the ratios $V_{2.2}/V_{\rm vir}$ and reduces average
$m_d$ values by nearly a factor of two.  This reduction improves the
agreement with observations by putting galaxies in halos of higher
mass and lower abundance.  The light--IMF model with AC lies midway
between the Kroupa--IMF models with and without AC.  It is marginally
consistent with the \citet{bell_etal03} mass function.

A third way to reconcile the TF and mass function constraints is to
change cosmological parameters, and thus the halo mass function.  We
find that lowering the matter density parameter to $\Omega_{\rm m}
\approx 0.15$ brings the prediction of Kroupa--IMF, AC model low
enough to match the knee of the observed mass function, and it also
brings the light--IMF, AC model into rough alignment with the
Kroupa--IMF, no AC model.  Changing the power spectrum normalization,
$\sigma_8$, on the other hand, has little impact on the predicted
$n(M_*)$ because the halo mass function is only weakly sensitive to
$\sigma_8$ in this mass range.

This discussion of the effects of changing cosmological parameters
ignores the potentially crucial impact of $\sigma_8$ on halo
concentrations.  \citet{bullock_etal01} present an analytic model for
predicting halo concentrations \citep[see also][]{kuhlen_etal05},
calibrated against their $\Omega_{\rm m}=0.3$, $\sigma_8=1.0$
$\Lambda$CDM simulation.  If we apply this model to the WMAP 3-year
cosmological parameters, we obtain concentrations lower by about 25\%
at fixed mass, mainly because of the lower value of $\sigma_8$.  In
the range $M = 10^{11} - 10^{13} \, h^{-1} \Msun$, the analytic model
prediction is well fit by a power law, $\bar{c}(M) \approx 9 \,
(M/10^{12}\, h^{-1} \Msun)^{-0.1}$.  We have not adopted this as our
standard relation for the calculations above because the $\sigma_8$
dependence of halo concentrations has not been thoroughly tested
against $N$-body simulations.  However, the preliminary tests that
have been carried out to date, and the tests of the predicted
redshift dependence of concentrations, suggest that the analytic model
is accurate (J. Bullock, private communication).

The last lines of Tables \ref{tab:tf} and \ref{tab:fp} list TF and FP
results for a model similar to our best-fit Kroupa--IMF, AC model with
$\bar{m}_d \propto \Sigma_*^{0.65}$, but using this lower $\bar{c}(M)$
relation.  Lower concentrations place a smaller fraction of the halo
mass within $2.2\, R_d$, and therefore the total halo mass must be
larger to reproduce the observed $V_{2.2}$.  The best-fit value of
$\bar{m}_{d,0}$ (defined at $M_* \, R_d^{-2} = 10^{9.2}\, \Msun\, {\rm
kpc}^{-2}$, see eq. [\ref{eq:sden}]) drops from 0.1 to 0.04.  For a
galaxy with the mean $\bar{c}(M)$ that lies on the mean $R_d - M_*$
relation, this change of normalization keeps the mass of the
adiabatically-contracted NFW halo within $2.2\, R_d$ approximately the
same as before.  Lowering $\bar{c}(M)$ and $\bar{m}_{d,0}$ yields a
slightly better fit to the TF relation, with weaker residual
anti-correlation but slightly worse slope and intercept, and a
noticeably better fit to the FP relation.  However, the improvement in
the predicted mass function is dramatic, as shown by the long-dashed
curve in Figure~\ref{fig:mf}, because the lower $m_d$ values put
galaxies in more massive, less abundant halos.  The low $\bar{c}(M)$,
AC model fares just slightly worse than the high $\bar{c}(M)$, no--AC
model in the mass function test.

% \citet{conroy_etal06} matched the number densities of galaxies above a
% given luminosity, $L$, with the halos above a given maximum circular
% velocity, $V_{\rm max}$, to construct a model linking galaxies with
% halos via the relation $n_g(>L) = n_h(>V_{\rm max})$.  This model
% successfully reproduced clustering properties of various galaxy
% samples from $z=0$ to $z \approx 4$.  At $z=0$, their model matches
% the luminosity function of SDSS galaxies in $r$-band
% \citep{blanton_etal03} with the abundance of halos in a cosmological
% $N$-body simulation.  We can compare the prediction of this model with
% our best-fit values for the dynamical mass-to-light ratio.  To do
% that, we have converted $V_{\rm max}$ to virial velocities, $V_{\rm
% vir}$, and then to virial masses, $M$, using mean halo concentrations,
% $\bar{c}(M)$.  Their resulting $M_{\rm vir}/L_r$ ratios are
% significantly above ours, on average by a factor 10, although their
% model values have scatter and some outliers overlap with our $M_{\rm
% vir}/L_r$ values.  Such a large discrepancy in the dynamical
% mass-to-light ratios is related to the inability of our models to fit
% the luminosity function (\S\ref{sec:lf}), while their model reproduces
% it by construction.

\begin{figure}[t]
\vspace{-1.1cm}
\centerline{\epsfysize3.5truein \epsffile{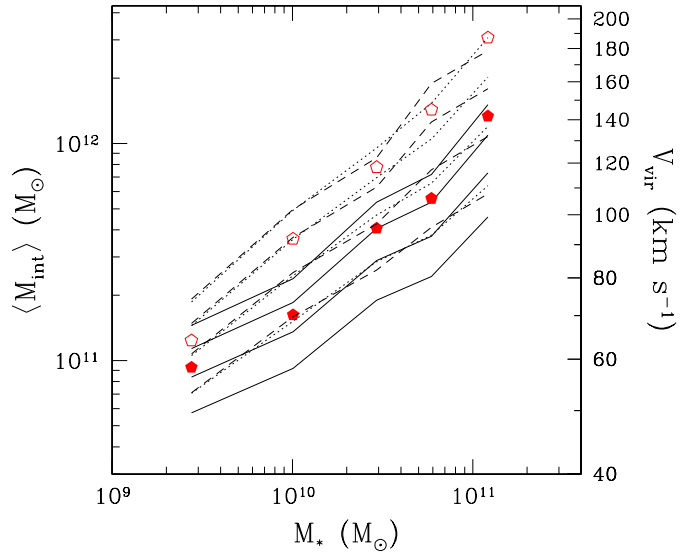}}
\vspace{-0.3cm}
\caption{Average mass enclosed within 50, 100, 200, and 400 kpc
  (bottom to top) of galaxy centers as a function of stellar mass for
  the best-fit models with $\bar{m}_d \propto \Sigma_*^{p_2}$, with AC
  ({\it solid lines}), without AC ({\it dotted lines}), and with AC
  and light stellar IMF ({\it dashed lines}).  Pentagons show the
  average halo virial mass for the standard IMF, with AC ({\it
  filled}) and without AC ({\it open}).}
  \label{fig:mvirms}
\end{figure}

Using our best-fit models, we can also make predictions for extended
mass distributions around galaxies like those in our sample.
Figure~\ref{fig:mvirms} shows the mean total mass enclosed within a
fixed distance from the galaxy center, for distances ranging from
$d=50$ kpc to $d=400$ kpc.  For each galaxy in our sample, we use one
realization of each model to calculate the interior mass $M_{\rm
int}(<d)$ as a sum of the observed stellar mass and the predicted dark
halo mass with the NFW profile.  The distribution of gas (if present)
is implicitly assumed to follow the distribution of dark matter.  At
these large distances, we ignore the effect of AC on the shape of the
NFW profile, although the model $m_d$ values are of course determined
taking AC into account at $2.2\, R_d$.  We average $M_{\rm int}(<d)$
in five bins of stellar mass containing approximately equal number of
galaxies, and plot the mean $M_{\rm int}(<d)$ against the average
stellar mass $\left<M_*\right>$ in the bin.  For the AC and no--AC
models, we also show the bin average of the virial mass (for critical
overdensity $\Delta = 101$), with the corresponding virial velocity
marked on the right axis.  The interior masses for the no--AC model
are higher than those of the AC model, typically by factors of
$1.5-2$, because of the lower $m_d$ values.  The dependence of $M_{\rm
int}(<d)$ on $M_*$ is shallower for the AC model because of the
stronger dependence of $\bar{m}_d$ on $\Sigma_*$, which puts lower
mass (and generally lower $\Sigma_*$) galaxies in more massive halos.
The predicted interior mass in the models with $\bar{m}_d \propto
M_*^{p_1}$ is similar to that in the corresponding models with
$\bar{m}_d \propto \Sigma_*^{p_2}$.  Predictions of the light--IMF
model are close to those of the no--AC model, since they have similar
values of $m_d$.  Predictions of the Kroupa--IMF model with AC and low
$\bar{c}(M)$ are also similar to those of the plotted no--AC model at
$d \le 100$ kpc, but they lie systematically above it at larger $d$
because the halos are more extended.

\begin{figure}[t]
\vspace{-0.5cm}
\centerline{\epsfysize3.5truein \epsffile{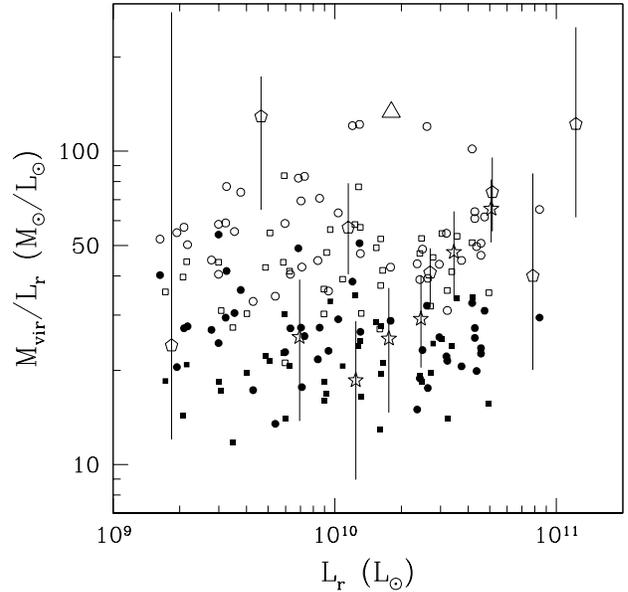}}
\vspace{-0.3cm}
\caption{Dynamical mass-to-light ratio in the $r$-band, $M_{\rm
  vir}/L_r$, for the best-fit AC models with $\bar{m}_d \propto
  \Sigma_*^{p_2}$, for Kroupa IMF ({\it filled squares and circles})
  and light IMF ({\it open squares and circles}).  In both cases,
  squares correspond to compact galaxies with $R_d < \bar{R}_d(M_*)$
  (eq. [\ref{eq:r-m}]), while circles correspond to extended galaxies
  with $R_d \ge \bar{R}_d(M_*)$.  Plotted luminosities $L_r$ are the
  observed values, not corrected for internal extinction.  Star
  symbols with error bars show the galaxy-galaxy weak lensing
  measurements from \protect\citet{hoekstra_etal05}, while open
  pentagons are from \protect\citet{mandelbaum_etal05}.  Open triangle
  shows an average measurement using satellite dynamics from Prada et
  al. (2006, in prep.).}
  \label{fig:mtol}
\end{figure}

These extended mass distributions can probed by galaxy-galaxy weak
lensing observations, which yield the mean total mass profiles after
stacking a large number of galaxies of similar luminosity or stellar
mass.  To compare with such measurements, we plot in
Figure~\ref{fig:mtol} the virial mass-to-light ratios in $r$-band,
$M_{\rm vir}/L_r$, predicted by the AC models with the Kroupa IMF and
the light IMF.  For each galaxy in our sample, we draw $m_d$ values
and concentrations from the $P(m_d | M_*, R_d)$ and $P(c)$
distributions 10 times and compute the average $M_{\rm vir}$, then
divide by the galaxy's observed $L_r$.  We follow the conventional
practice of lensing studies in using $L_r$ {\it not} corrected for
internal extinction (which differs from standard practice in TF
studies).  Pentagonal points with error bars show virial mass-to-light
ratios of late-type galaxies estimated by \citet{mandelbaum_etal05},
based on galaxy-galaxy lensing measurements in the SDSS.  These
ratios, ranging from $\sim 40 - 130 \, \Msun/\Lsun$, are consistent
with the light--IMF model but too high for the AC, Kroupa--IMF model.
Star symbols with error bars show the galaxy-galaxy lensing results of
\citet{hoekstra_etal05} using the Red-Sequence Cluster Survey, which
are somewhat lower and overlap with the predictions of both models.
Results for the Kroupa--IMF models with no AC and with AC but low
$\bar{c}(M)$, not shown in the Figure, are similar to those of the
light--IMF model with AC.

An independent measure of the dynamical mass at large distances from
the galaxy center is provided by the dynamics of satellite galaxies.
Using a large sample of SDSS galaxies, Prada et al. (2006, in prep.)
estimate $M_{\rm vir} = 2.4\times 10^{12}\, \Msun$ for the average
luminosity $L_r = 1.8\times 10^{10}\, \Lsun$ (open triangle in
Figure~\ref{fig:mtol}).  After converting luminosity to stellar mass,
this corresponds to $m_d = 0.028$, below our best-fit values at this
$M_*$ ($\bar{m}_d = 0.11$ and 0.044, for the Kroupa--IMF and
light--IMF models with AC, respectively, and $\bar{m}_d = 0.042$ for
the no--AC model).  Note that this estimated mass fraction applies to
an average of galaxies of all types, which could plausibly lie below
the mass fraction of disk-dominated galaxies in our sample.  Stellar
spheroids (bulges and elliptical galaxies) have higher {\it stellar}
mass-to-light ratios than younger disk systems, so they have higher
$M_{\rm vir}/L$ even for the same $M_*/M_{\rm vir}$.
\citet{mandelbaum_etal05} find an average $M_{\rm vir}/L_r$ of 140 for
early-type galaxies vs. 53 for late types in the luminosity range $L_r
= 10^{10} - 10^{11}\, \Lsun$.  Furthermore, recent studies suggest
that bulge-dominated galaxies have lower values of $M_*/M_{\rm vir}$.
For example, \citet{humphrey_etal06} find a stellar fraction of 0.044
for an X-ray model of an elliptical galaxy NGC 720, while
\citet{hoekstra_etal05} find on the average 2.5 times lower
$M_*/M_{\rm vir}$ for the red ($B-V > 0.8$) galaxies than the blue
galaxies.  Galaxies composed of a mixture of the disk and spheroidal
components would have lower average $M_*/M_{\rm vir}$ than pure
disk-dominated galaxies, and this difference may contribute to the gap
between our AC models and the Prada et al. (2006) measurement.

Available cosmological gasdynamics simulations of disk-dominated
galaxies, which automatically include the halo response to baryon
condensation, predict $m_d$ values in the same general range as our
models.  In a realization of 19 Milky Way-type galaxies,
\citet{sommer-larsen06} finds a mean baryon mass fraction $m_d = 0.08$
at $V_c = 200$ km s$^{-1}$.  The TF relation for this simulated sample
is roughly consistent with our observed relation
\citep{portinari_sommer-larsen06}, though the sample is too small to
investigate residual correlations.  For a single galaxy studied with
ultra-high resolution, \citet{robertson_etal04} also find $m_d \approx
0.08$, while for a sample of three disk galaxies
\citet{governato_etal06} find values of $m_d$ between 0.04 and 0.08.

All of our best-fit models introduce a correlation of $\bar{m}_d$ with
$\Sigma_*$, which counteracts the tendency of disk gravity to produce
higher rotation speeds in more compact galaxies.  For the model points
in Figure~\ref{fig:mtol}, circles and squares respectively shows
results for galaxies above and below the mean $R_d$ at a given $M_*$
(eq. [\ref{eq:r-m}]).  Larger $R_d$ galaxies have systematically
higher $M_{\rm vir}/L_r$ because they reside in more massive halos.
If we divide the sample into three equal-size subsamples containing
the smallest, intermediate, and largest galaxies for their $M_*$, the
mean Kroupa--IMF $M_{\rm vir}/L_r$ values are 20, 23, and 28,
respectively, and light--IMF $M_{\rm vir}/L_r$ values are 42, 46, and
60, respectively.  Galaxy-galaxy lensing analyses could test our
proposed explanation of the weak correlation between TF residuals and
radius, by detecting or ruling out this difference, roughly 40\%
between the smallest and largest 1/3 of the population.

Measurements of the rotation speed at distances larger than $2.2 \,
R_d$ may also help to differentiate among our proposed best-fit
models.  High density galaxies in our models have relatively smaller
dark matter halos and, therefore, more steeply falling rotation curves
at large distances than low density galaxies.  As an example, we
consider hypothetical measurements, $V_{3.2}$, of the rotation
velocity at $3.2 \, R_d$.  For the AC, Kroupa--IMF model, the average
ratio $V_{3.2}/V_{2.2}$ drops from 1.00 for the lowest-density
galaxies in our sample ($M_* \, R_d^{-2} \approx 2\times 10^8\,
\Msun\, {\rm kpc}^{-2}$) to 0.97 for the highest-density galaxies
($M_* \, R_d^{-2} \approx 7\times 10^9\, \Msun\, {\rm kpc}^{-2}$).
For the light--IMF and low $\bar{c}(M)$ models, $V_{3.2}/V_{2.2}$
varies from 1.02 to 0.99 in the same density range.  In contrast, for
the no--AC model the ratio $V_{3.2}/V_{2.2}$ is always in the range
$1.00 - 1.03$, indicating that the rotation curve is still rising at
$3.2 \, R_d$.  Measurements at $4\, R_d$ would show a larger
difference.  In the no--AC model, the ratio $V_{4}/V_{2.2}$ is still
close to unity, while in the AC, Kroupa--IMF model it drops from 0.99
to 0.93, for the density range of our sample.  This variation
corresponds to up to 15 km s$^{-1}$ difference in the rotation speed,
which could be detectable in a sample of extended, high-precision
rotation curves.

% Throughout this paper, we have adopted the NFW profile for all halos,
% with the range of concentration parameters predicted by $N$-body
% simulations.  Although recent high-resolution studies of halo profiles
% suggest possible flattening of the central density slope
% \citep[e.g.,][]{navarro_etal04}, this should have minimal impact on
% the mass within $2.2\, R_d$, and adiabatic contraction reduces memory
% of the initial halo profile at these scales in any case.  However,
% changes to the initial halo profiles or to the baryon-dark matter
% interaction that substantially altered the fraction of halo mass
% within $2.2\, R_d$ would change our best-fit $m_d$ values and,
% consequently, our predictions for galaxy mass function and extended
% mass distributions.  In general, changes that reduce the central
% density of halos lead to lower $m_d$ ratios and better agreement with
% the mass function, weak lensing, and satellite dynamics data.  On the
% opposite end, models with insufficient dark matter within $2.2\, R_d$
% will run afoul of the TF residual constraints as a result of overly
% dominant stellar disks.

\section{Conclusions}

We have constrained the structural properties of dark matter halos of
disk-dominated galaxies using a well-defined sample of 81 late-type
galaxies with H$\alpha$ rotation curves, selected from the SDSS
redshift survey \citep{pizagno_etal05}.  We model the Tully-Fisher and
fundamental plane relations constructed from the galaxy stellar mass,
$M_*$, disk scale length, $R_d$, and optical H$\alpha$ rotation
velocity at $2.2 \, R_d$, $V_{2.2}$.  The stellar mass is determined
from the $i$-band luminosity and a stellar mass-to-light ratio
estimated from the galaxy's $g-r$ color.  The normalization of these
stellar masses depends on the IMF, for which we generally adopt the
form of \citet{kroupa01}.  The maximum likelihood fit to the TF
relation for our sample is $\log{V_{2.2}} = 0.29 (\log{M_*} - 10.5) +
2.23$.  The FP relation is $\log{V_{2.2}} = 0.27 (\log{M_*} - 10.5) +
0.085 (\log{R_d} - 0.65) + 2.23$.  Residuals from the TF relation are
weakly, and positively, correlated with residuals from the $R_d - M_*$
relation.  The weakness of this correlation, and the small $R_d$
coefficient in the FP relation, imply that the TF relation is a nearly
edge-on view of the fundamental plane of disk galaxies.

We define models of the disk galaxy population by the probability
distribution function $P(m_d | M_*,R_d)$ of the disk stellar mass
fraction $m_d = M_*/M_{\rm vir}$, where $M_{\rm vir}$ is the virial
mass of the host dark matter halo, assumed to have an NFW profile with
the concentration predicted by $N$-body studies.  For each sample
galaxy, we draw $m_d$ from the model distribution, then find the disk
angular momentum parameter $\lambda_d$ required to reproduce the
galaxy's observed disk scale length.  Our standard calculations
include the impact of adiabatic contraction (AC) on the halo profile
\citep[e.g.,][]{blumenthal_etal86, gnedin_etal04}, but we also consider
models without AC.

Models with a constant value of $m_d$ require $m_d \approx 0.08$ with
AC ($m_d \approx 0.04$ without AC) to reproduce the TF intercept and
slope.  These models under-predict the scatter in the TF relation,
even when we include the full range of concentration scatter found in
$N$-body simulations of dark matter halos.  More importantly, they
predict a strong anti-correlation of TF and $R_d - M_*$ residuals,
with more compact galaxies rotating faster at fixed $M_*$.
Equivalently, they predict an $R_d$ coefficient in the FP relation
that has the wrong sign and is incompatible with the observed value.
Adding random scatter to $V_{2.2}$ values to match the observed
intrinsic scatter does not resolve this conflict with the observed
residual correlations.  The conflicts are more severe for the models
with AC because of the stronger impact of disk gravity.  The AC models
also predict a TF slope that is slightly too steep.

We conclude that a scatter of $m_d$ values, with dispersion in
$\ln{m_d}$ of $\sigma_m \approx 0.25$, is required to reproduce the
observed scatter of the TF relation, and that models with AC require
higher disk mass fractions in higher mass galaxies to reproduce the
observed slope.  These findings confirm the more qualitative arguments
of \citet{pizagno_etal05} based on the same data set.  We can obtain
consistency with the weak TF residual correlation and the small $R_d$
FP coefficient by tying the mean disk mass fraction to the stellar
surface density.  The best-fit model with AC has $\bar{m}_d \propto
\Sigma_*^{0.65}$ and a mean fraction $\bar{m}_d = 0.1$ at $M_* \,
R_d^{-2} = 10^{9.2}\, \Msun\, {\rm kpc}^{-2}$.  Without AC, the
best-fit model has a shallower slope, 0.2 instead of 0.65, and a lower
normalization, $\bar{m}_d = 0.04$ at $10^{9.2}\, \Msun\, {\rm
kpc}^{-2}$.  The $\bar{m}_d - \Sigma_*$ correlation counteracts the
effect of stronger disk gravity in more compact galaxies.  It produces
a major improvement for the AC model, and a modest improvement for the
no--AC model.

The derived $\lambda_d$ distributions for these two models are
significantly narrower than the $\lambda$ distribution for dark matter
halos in $N$-body simulations: $\sigma_{\lambda d} \approx 0.25$ (AC)
or 0.39 (no AC) vs. $\sigma_\lambda \approx 0.56$.  The mean
$\bar{\lambda}_d$ depends on the typical $m_d$ values, which determine
halo virial radii: we find $\bar{\lambda}_d \approx 0.054$ (AC) or
$\bar{\lambda}_d \approx 0.036$ (no AC) vs.  $\bar{\lambda} \approx
0.045$ for $N$-body halos.  The systematic difference between the
$\lambda_d$ and $\lambda$ distributions implies either that disk
baryons have systematically different angular momentum from the dark
matter in their host halo or that only a subset of halos form
disk-dominated galaxies like those in our sample.

We consider the $\bar{m}_d - \Sigma_*$ correlation a physically
plausible mechanism for bringing theoretically motivated disk
formation models, which include adiabatic contraction, into agreement
with the slope and residual correlation of the observed TF relation.
However, our best-fit AC model suffers the long-standing problem of
over-producing the galaxy stellar mass function given the halo
population of a $\Lambda$CDM cosmological model with observationally
favored parameter values and the concentration-mass relation of
\citet{bullock_etal01}.  It also predicts virial mass to $r$-band
light ratios lower than some recent estimates from galaxy-galaxy
lensing and satellite dynamics.  Lowering $\Omega_{\rm m}$, and thus
the halo mass function, can ameliorate the abundance problem, but the
required value, $\Omega_{\rm m} \approx 0.15$, seems implausibly low
and the mass-to-light ratio problem remains in any case.  Eliminating
AC can resolve both problems, at the expense of dropping a well-tested
ingredient of galaxy formation theory.

We suggest an alternative solution: adopting a ``light'' stellar IMF
with $M_*/L$ lower than the \citet{kroupa01} IMF by about 0.15 dex.
This model yields acceptable residual correlations with a modest trend
$\bar{m}_d \propto \Sigma_*^{0.4}$, predicts virial mass-to-light
ratios in accord with recent measurements, and reduces but does not
entirely remove the tension with the galaxy baryon mass function.  We
do not know of other observational evidence to support such a light
IMF, but it is perhaps within the uncertainties of existing
observational constraints.  Systematic uncertainties in the
\citet{bell_etal03} population synthesis models, which we use to
compute $M_*/L$ from $g-r$ color, could account for some of the
required change even without altering the IMF itself.
We also show that the lower $\bar{c}(M)$ relation predicted for
$\sigma_8 = 0.74$ by Bullock et al.'s (2001b) analytic model relieves
most of the tension between the TF relation and the mass function and
$M_{\rm vir}/L_r$ constraints, even in the case of a Kroupa IMF.  A
full assessment of this solution awaits better numerical confirmation
of the analytic model and better observational constraints on
$\sigma_8$.  Other changes to halo profiles that reduce the fraction
of mass at small radii would have similar effects.

Our observational findings are consistent with those of
\citet{dutton_etal06}, which is reassuring given the different
characteristics of the data sets.  Our theoretical inferences are
consistent but not identical.  We are more reluctant than
\citet{dutton_etal06} to abandon adiabatic contraction, given the
numerous simulations in which it occurs and the absence of any
simulations in which it does not.  The $\bar{m}_d - \Sigma_*$
correlation can remove the $\Delta R_d - \Delta V_{2.2}$
anti-correlation that otherwise plagues models with AC, though it
leaves problems with the baryon mass function and weak lensing
mass-to-light ratios.  The addition of a light IMF or the adoption of
lower halo concentrations largely resolve these difficulties as well,
and we consider some combination of these effects to be a plausible
resolution of the observational tensions.  Modeling of rotation curve
data at larger and smaller separations and galaxy-galaxy lensing
measurements for more finely divided galaxy sub-classes may provide
further insight that could help discriminate among these solutions.
The fact that disk formation models based on the most straightforward
theoretical and observational inputs do not reproduce all aspects of
the current data suggests that interesting surprises may still lie
ahead.

\acknowledgements 
We thank Stacy McGaugh and Marla Geha for discussions and comparisons
of the observed samples, Jaiyul Yoo for calculating halo
concentrations in the new WMAP cosmology using \citet{bullock_etal01}
model, and James Bullock for useful insights on halo concentrations.
% Charlie Conroy for supplying the catalog of his $L-V_{\rm max}$ model.
This work was supported in part by NASA ATP grant NNG04GK68G
and NSF Grant AST-0407125. 

This paper is based on public data of the Sloan Digital Sky Survey
(SDSS).  Funding for the SDSS and SDSS-II has been provided by the
Alfred P. Sloan Foundation, the Participating Institutions, the
National Science Foundation, the U.S. Department of Energy, the
National Aeronautics and Space Administration, the Japanese
Monbukagakusho, the Max Planck Society, and the Higher Education
Funding Council for England.  The Participating Institutions are
listed at the official SDSS web site, http://www.sdss.org/.

\begin{appendix}
\section{Multidimensional Linear Fit using Maximum Likelihood}

We have developed the following procedure for a multidimensional
linear fit to data with observational errors in all variables,
allowing for intrinsic Gaussian scatter.  Each data point (galaxy) is
represented by $n+1$ parameters, $n$-vector $\bx$ and scalar $y$, and
their associated measurement errors, $\bsigma_{x}$ and $\sigma_{y}$.
We label each galaxy by subscript $i$ and fit a linear relation of the
form
\begin{equation}
  y = \ba (\bx - \bx_0) + b + N(0,\sigma),
\end{equation}
where $\ba$ is also an $n$-dimensional vector and $\bx_0$ is an
arbitrary offset.  The term $N(0,\sigma)$ accounts for
normally-distributed scatter that is intrinsic to the studied relation
and is not accounted for by measurement errors.  In the case of the TF
relation, such scatter could be due to the variations of the
mass-to-light ratio among the galaxies, scatter in the properties of
their dark matter halos, etc.  We define the likelihood function
\begin{equation}
  2 \ln{\cal L} =
  - \sum_i \ln{(\sigma^2 + \sigma_{y,i}^2 + \ba^2\bsigma_{x,i}^2)}
  - \sum_i {[\ba(\bx_i - \bx_0) + b - y_i]^2 \over 
          \sigma^2 + \sigma_{y,i}^2 + \ba^2\bsigma_{x,i}^2}
  + \,{\rm const}
\end{equation}
and find its maximum using Powell's direction set method \citep{NR}.
The direction search is done as a function of $n+1$ variables ($\ba$
and $\sigma$), since the value of the intercept $b$ can then be
calculated analytically:
\begin{equation}
  b = {\sum_i w_i [y_i - \ba(\bx_i - \bx_0)] \over \sum_i w_i}, \qquad 
      w_i \equiv {1 \over \sigma^2 + \sigma_{y,i}^2 + \ba^2\bsigma_{x,i}^2}.
\end{equation}

We estimate the errors of the best-fit parameters ($\bdelta \ba$,
$\Delta b$, $\Delta \sigma$) by bootstrapping the data.  The error
estimates are taken to be the standard deviation of the corresponding
distributions of the best-fit parameters.  We found empirically that
of the order 1000 bootstrap realizations are needed in order to obtain
robust estimates of the errors, accurate to 3\%.

The offset $\bx_0$ can be chosen so as to minimize the uncertainty
$\Delta b$ and eliminate correlation between $\bdelta \ba$ and $\Delta
b$.  The slope of the linear fit ($\ba$) and its error ($\bdelta \ba$)
do not depend on the choice of $\bx_0$ because its variation can be
absorbed in the value of the intercept $b$.  The intrinsic scatter and
its error are also unaffected by the choice of $\bx_0$.  The error of
the intercept $b$, however, does depend on $\bx_0$ and can be
minimized by the appropriate choice of the offset.  We use Brent's
parabolic interpolation method (if $n=1$) or downhill simplex method
(if $n>1$) to find the minimum of $\Delta b$.

Computer code that calculates the best-fit parameters and their error
estimates is available from the authors upon request.  An independent
derivation of this maximum likelihood method and further discussion
can be found in \citet{dagostini05}.
\end{appendix}

\begin{table}[t]
\begin{center}
\caption{\sc Tully-Fisher fit parameters
   \label{tab:tf}} %\\[2mm]
\begin{tabular}{llcccllllc}
\tableline\tableline\\
\multicolumn{1}{c}{$\bar{m}_d$} &
\multicolumn{1}{c}{$p_i$} &
\multicolumn{1}{c}{$\sigma_m$} &
\multicolumn{1}{c}{$\sigma_c$} &
\multicolumn{1}{c}{AC} &
\multicolumn{1}{c}{$a$} &
\multicolumn{1}{c}{$b$} &
\multicolumn{1}{c}{$\sigma$} &
\multicolumn{1}{c}{$r$} &
\multicolumn{1}{c}{$\chi^2$}
\\[2mm] \tableline\\
data &&   &    &   & $0.291\pm0.013$ & $2.231\pm0.006$ & $0.049\pm0.007$ & $0.23\pm0.14$ \\
%err$\times2$ &&& & & $0.290\pm0.013$ & $2.235\pm0.007$ & $0.027\pm0.012$ \\[2mm]
\tableline
0.02 &     & 0    & 0    & N & 0.280 & 2.265 & 0.009 & $-0.02$ &  67 \\
0.02 &     & 0    & 0    & Y & 0.292 & 2.306 & 0.014 & $-0.20$ & 171 \\
0.04 &     & 0    & 0    & N & 0.295 & 2.232 & 0.016 & $-0.26$ &  36 \\
0.04 &     & 0    & 0    & Y & 0.307 & 2.275 & 0.023 & $-0.41$ &  84 \\
0.08 &     & 0    & 0    & N & 0.311 & 2.196 & 0.027 & $-0.48$ &  68 \\
0.08 &     & 0    & 0    & Y & 0.322 & 2.240 & 0.034 & $-0.57$ &  46 \\[2mm]

% 0.04 & +$\sigma$ & 0 & 0 & N & 0.301 & 2.231 & 0.050 & $-0.25$ & 3.1 \\
% 0.04 & +$\sigma$ & 0 & 0 & Y & 0.313 & 2.275 & 0.050 & $-0.41$ & 17 \\
% 0.08 & +$\sigma$ & 0 & 0 & N & 0.317 & 2.195 & 0.050 & $-0.49$ & 15 \\
% 0.08 & +$\sigma$ & 0 & 0 & Y & 0.329 & 2.239 & 0.050 & $-0.64$ & 12 \\[2mm]

0.04 &     & 0    & 0.14 & N & 0.297 & 2.232 & 0.041 & $-0.22$ &  13 \\
0.04 &     & 0    & 0.14 & Y & 0.308 & 2.276 & 0.039 & $-0.36$ &  70 \\
0.08 &     & 0    & 0.14 & N & 0.313 & 2.196 & 0.040 & $-0.43$ &  55 \\
0.08 &     & 0    & 0.14 & Y & 0.323 & 2.240 & 0.041 & $-0.53$ &  40 \\[2mm]

0.04 & +$\sigma$ & 0 & 0.14 & N & 0.303 & 2.233 & 0.049 & $-0.27$ & 14 \\
0.04 & +$\sigma$ & 0 & 0.14 & Y & 0.314 & 2.276 & 0.049 & $-0.42$ & 72 \\
0.08 & +$\sigma$ & 0 & 0.14 & N & 0.319 & 2.196 & 0.049 & $-0.51$ & 60 \\
0.08 & +$\sigma$ & 0 & 0.14 & Y & 0.329 & 2.240 & 0.049 & $-0.65$ & 52 \\[2mm]

0.04 &     & 0.22 & 0.14 & N & 0.298 & 2.233 & 0.049 & $-0.20$ & 10 \\
0.08 &     & 0.27 & 0.14 & Y & 0.320 & 2.249 & 0.049 & $-0.43$ & 35 \\[2mm]
\tableline
\multicolumn{10}{c}{$\bar{m}_d \propto M_*^{p_1}$}\\[2mm]
0.04 & 0.1 & 0.25 & 0.14 & N & 0.283 & 2.234 & 0.051 & $-0.17$ & 8.7 \\
0.1  & 0.3 & 0.21 & 0.14 & Y & 0.291 & 2.239 & 0.046 & $-0.45$ & 25 \\[2mm]
\tableline
\multicolumn{10}{c}{$\bar{m}_d \propto \Sigma_*^{p_2}$}\\[2mm]
0.04 & 0.2  & 0.25 & 0.14 & N & 0.282 & 2.234 & 0.051 & $-0.07$ & 5.4 \\
0.1  & 0.65 & 0.2  & 0.14 & Y & 0.281 & 2.243 & 0.041 & $-0.18$ & 14 \\
0.04 & 0.4  & 0.24 & 0.14 & N & 0.266 & 2.236 & 0.051 & $+0.02$ & 6.9 \\
0.1  & 0.9  & 0.2  & 0.14 & Y & 0.264 & 2.247 & 0.044 & $-0.08$ & 16 \\[2mm]
\tableline
\multicolumn{10}{c}{$\bar{m}_d \propto \Sigma_*^{p_2}$, light IMF}\\[2mm]
0.04 & 0.4  & 0.25 & 0.14 & Y & 0.273 & 2.275 & 0.051 & $+0.01$ & 4.6 \\[2mm]
\tableline
\multicolumn{10}{c}{$\bar{m}_d \propto \Sigma_*^{p_2}$, low $\bar{c}(M)$}\\[2mm]
0.04 & 0.65 & 0.2  & 0.14 & Y & 0.279 & 2.246 & 0.041 & $-0.06$ & 12 \\[2mm]
\tableline
\end{tabular}
\end{center}
{\small Note -- The intercept $b$ is for the offset $\log{M_{*0}} =
  10.5$.  Lines with ``+$\sigma$'' correspond to models with added
  measurement errors to match the observed scatter.  In the ninth
  column, $r$ is Pearson's linear correlation coefficient.  We
  estimate its error by bootstrapping the observed sample.  We find
  that $\Delta r$ depends mainly on the number of objects in the
  sample, $N$, and approximately equals $N^{-1/2}$.  For the light
  IMF, the observed intercept $b = 2.231 + 0.15\times 0.291 = 2.275$.
  \vspace{0.5cm}}
\end{table}

\begin{table}[t]
\begin{center}
\caption{\sc Fundamental plane fit parameters
  \label{tab:fp}} %\\[2mm]
\begin{tabular}{llcccllllc}
\tableline\tableline\\
\multicolumn{1}{c}{$\bar{m}_d$} &
\multicolumn{1}{c}{$p_i$} &
\multicolumn{1}{c}{$\sigma_m$} &
\multicolumn{1}{c}{$\sigma_c$} &
\multicolumn{1}{c}{AC} &
\multicolumn{1}{c}{$a_1$} &
\multicolumn{1}{c}{$a_2$} &
\multicolumn{1}{c}{$b$} &
\multicolumn{1}{c}{$\sigma$} &
\multicolumn{1}{c}{$\chi^2$}
\\[2mm] \tableline\\
data &&&&& $0.271\pm0.017$ & $0.085\pm0.054$ & $2.233\pm0.006$ & $0.048\pm0.006$\\
%err$\times2$ &&&&& $0.271\pm0.018$ & $0.076\pm0.051$ & $2.211\pm0.006$ & $0.029\pm0.012$\\[2mm]
\tableline
0.02 &     & 0    & 0    & N & 0.280 & $-0.002$ & 2.265 & 0.009 &  69 \\
0.02 &     & 0    & 0    & Y & 0.305 & $-0.054$ & 2.305 & 0.008 & 184 \\
0.04 &     & 0    & 0    & N & 0.312 & $-0.075$ & 2.230 & 0.008 &  55 \\
0.04 &     & 0    & 0    & Y & 0.336 & $-0.125$ & 2.272 & 0.007 & 111 \\
0.08 &     & 0    & 0    & N & 0.347 & $-0.154$ & 2.192 & 0.008 & 125 \\
0.08 &     & 0    & 0    & Y & 0.369 & $-0.200$ & 2.234 & 0.007 & 104 \\[2mm]

% 0.04 & +$\sigma$ & 0 & 0 & N & 0.321 & $-0.084$ & 2.229 & 0.048 & 4.7 \\
% 0.04 & +$\sigma$ & 0 & 0 & Y & 0.346 & $-0.138$ & 2.271 & 0.045 & 19 \\
% 0.08 & +$\sigma$ & 0 & 0 & N & 0.358 & $-0.167$ & 2.191 & 0.043 & 23 \\
% 0.08 & +$\sigma$ & 0 & 0 & Y & 0.381 & $-0.217$ & 2.234 & 0.038 & 19 \\[2mm]

0.04 &     & 0    & 0.14 & N & 0.315 & $-0.076$ & 2.230 & 0.038 & 18 \\
0.04 &     & 0    & 0.14 & Y & 0.338 & $-0.125$ & 2.272 & 0.032 & 77 \\
0.08 &     & 0    & 0.14 & N & 0.349 & $-0.155$ & 2.192 & 0.030 & 93 \\
0.08 &     & 0    & 0.14 & Y & 0.370 & $-0.201$ & 2.235 & 0.024 & 76 \\[2mm]

0.04 & +$\sigma$ & 0 & 0.14 & N & 0.325 & $-0.093$ & 2.231 & 0.048 & 21 \\
0.04 & +$\sigma$ & 0 & 0.14 & Y & 0.349 & $-0.143$ & 2.273 & 0.045 & 80 \\
0.08 & +$\sigma$ & 0 & 0.14 & N & 0.361 & $-0.174$ & 2.192 & 0.043 & 96 \\
0.08 & +$\sigma$ & 0 & 0.14 & Y & 0.382 & $-0.220$ & 2.235 & 0.038 & 76 \\[2mm]

0.04 &     & 0.22 & 0.14 & N & 0.315 & $-0.075$ & 2.230 & 0.047 & 16 \\
0.08 &     & 0.27 & 0.14 & Y & 0.361 & $-0.172$ & 2.244 & 0.038 & 56 \\[2mm]
\tableline
\multicolumn{10}{c}{$\bar{m}_d \propto M_*^{p_1}$}\\[2mm]
0.04 & 0.1 & 0.25 & 0.14 & N & 0.298 & $-0.064$ & 2.234 & 0.051 & 10 \\
0.1  & 0.3 & 0.21 & 0.14 & Y & 0.332 & $-0.177$ & 2.235 & 0.035 & 41 \\[2mm]
% 0.04 & 0.3  & 0.24 & 0.14 & N & 0.267 & $-0.048$ & 2.234 & 0.051 & 6.4 \\
% 0.1  & 0.7  & 0.2  & 0.14 & Y & 0.274 & $-0.136$ & 2.244 & 0.040 & 22 \\[2mm]
\tableline
\multicolumn{10}{c}{$\bar{m}_d \propto \Sigma_*^{p_2}$}\\[2mm]
0.04 & 0.2  & 0.25 & 0.14 & N & 0.287 & $-0.018$ & 2.234 & 0.051 & 4.6 \\
0.1  & 0.65 & 0.2  & 0.14 & Y & 0.296 & $-0.066$ & 2.241 & 0.039 & 14 \\
0.04 & 0.4  & 0.24 & 0.14 & N & 0.261 & $+0.022$ & 2.237 & 0.051 & 2.2 \\
0.1  & 0.9  & 0.2  & 0.14 & Y & 0.270 & $-0.022$ & 2.247 & 0.043 & 9.6 \\[2mm]
\tableline
\multicolumn{10}{c}{$\bar{m}_d \propto \Sigma_*^{p_2}$, light IMF}\\[2mm]
0.04 & 0.4  & 0.25 & 0.14 & Y & 0.268 & $+0.018$ & 2.275 & 0.051 & 1.8 \\[2mm]
\tableline
\multicolumn{10}{c}{$\bar{m}_d \propto \Sigma_*^{p_2}$, low $\bar{c}(M)$}\\[2mm]
0.04 & 0.65 & 0.2  & 0.14 & Y & 0.283 & $-0.017$ & 2.246 & 0.041 & 9.8 \\[2mm]
\tableline
\end{tabular}
\end{center}
{\small Note -- The intercept $b$ is for the offsets $\log{M_{*0}} =
  10.5$ and $\log{R_{d0}} = 0.65$.  For the light IMF, the observed
  intercept $b = 2.275$.
  \vspace{0.5cm}}
\end{table}

\bibliography{tf}
\end{document}